\PassOptionsToPackage{tbtags}{amsmath}
\PassOptionsToPackage{dvipsnames}{color}

\documentclass{jfm}

\usepackage{graphicx}
\usepackage{float}
\usepackage{newtxtext}
\usepackage{newtxmath}
\usepackage{natbib}
\usepackage{hyperref}
\hypersetup{
 colorlinks = true,
 urlcolor  = blue,
 citecolor = black,
 }
\DeclareGraphicsExtensions{.pdf,.eps,.png}

\usepackage[normalem]{ulem}

\newcommand{\add}[1]{\bgroup\color[named]{OrangeRed}#1\egroup}

\newcommand{\com}[1]{\bgroup\color[named]{JungleGreen}[#1]\egroup}

\newcommand{\AKcorrect}[1]{\bgroup\color[named]{RoyalBlue}#1\egroup}

\newlength{\figwidth}
\useAMSsubequations
\let\AMSsubequations=\subequations
\let\endAMSsubequations=\endsubequations
\renewenvironment{subequations}
 {\AMSsubequations
  \renewcommand{\theequation}{\theparentequation\textit{\alph{equation}}}}
 {\endAMSsubequations}

\hypersetup{pdfpagemode={UseOutlines},
bookmarksopen=true,
bookmarksopenlevel=0,
hypertexnames=false,
colorlinks=true, 
citecolor=blue, 
linkcolor=red, 
urlcolor=black, 
pdfstartview={FitV},
unicode,
breaklinks=true,
}
\usepackage{graphicx,amssymb,amsmath}
\usepackage{grffile}
\usepackage{color}
\usepackage{array}
\usepackage{hhline}

\usepackage{ulem}

\usepackage{siunitx}
\shorttitle{Vertical convection regimes}
\title{Vertical convection regimes in a two-dimensional {} rectangular cavity: Prandtl and aspect ratio dependance}
\author{Arman Khoubani\aff{1},
Ashwin Vishnu Mohanan\aff{2},
Pierre Augier\aff{1} \\
\and Jan-Bert Fl\'or\aff{1}
\corresp{\email{jan-bert.flor@cnrs.fr}}}

\affiliation{%
\aff{1}%
 Laboratoire des \'Ecoulements G\'eophysiques et Industriels,
 Universit\'e~Grenoble~Alpes, CNRS, Grenoble~INP,
 38000~Grenoble, France
\aff{2}%
 Swedish Meteorological and Hydrological Institute,\\
 Norrk{\"o}ping, Sweden.}

\shortauthor{A. Khoubani, A.V. Mohanan, P. Augier and J.B. Flór}

\begin{document}

\maketitle

\begin{abstract}
Vertical convection is the fluid motion that is induced by the heating and cooling of
two opposed vertical boundaries of a rectangular cavity  \citep[see e.g][]{Wang2021}.
We consider the linear stability of the steady two-dimensional flow reached at Rayleigh
numbers of O($10^8$).

As a function of the Prandtl number, $Pr$, and the height-to-width aspect ratio of the
domain, $A$, the base flow of each case is computed numerically and linear simulations
are used to obtain the properties of the leading linear instability mode. Flow regimes
depend on the presence of a circulation in the entire cavity, detachment of the thermal
layer from the boundary or the corner regions, and on the oscillation frequency
relative to the natural frequency of oscillation in the stably temperature-stratified
interior, allowing for the presence of internal waves or not. Accordingly the regime is
called slow or fast, respectively. Either the global circulation or internal waves in
the interior may couple the top and bottom buoyancy currents, while their absence
implies asymmetry in their perturbation amplitude.

Six flow regimes are found in the range of $0.1 \leq Pr \leq 4$ and $0.5 \leq A \leq
2$. For $Pr \lessapprox 0.4 $ and $A>1$ the base flow is driven by a large circulation
in the entire cavity. For $Pr \gtrapprox 0.7$ the thermal boundary layers are thin and
the instability is driven by the motion along the wall and the detached boundary layer.
A transition between these regimes is marked by a dramatic change in oscillation
frequency at $Pr = 0.55 \pm0.15$ and $A <2$.
\end{abstract}


\maketitle


\section{Introduction}

Over the past six decades, vertical convection has attracted significant interest due
to its wide range of applications in industry, the environment as well as in
geophysics. Circulation patterns and instabilities that may arise due to vertical heat
transport along hot or cold isothermal boundaries are relevant in view of the transport
of heat  \citep[see e.g.][]{Miroshnichenko2018}.  In the idealised case of a
rectangular cavity, the typical flow evolution is that, after turning on the heat
forcing above its critical value for convection, an upward motion arises at the heated
boundary and a downward motion at the cooled boundary, while stratification develops
progressively in the interior \citep{Gill1966}. When these motions reach the two
horizontal adiabatic boundaries, they turn into horizontal buoyancy currents. The flow
pattens in this cavity and related instabilities are determined by the Rayleigh number,
the Prandtl number, and the aspect ratio defined respectively as
\begin{equation}
Ra = \frac{g \beta \Delta T H^{3} }{\nu \kappa} \;\; Pr =\frac{\nu }{ \kappa}\;\; \mbox{ and }\;\; A=\frac{H}{W},
\end{equation}
with $\beta$ the thermal expansion coefficient, $\nu$ the dynamic viscosity, $\kappa$
the thermal diffusivity, and $H$ and $W$ the height and width of the cavity, $\Delta T$
the horizontal temperature difference in the cavity, and $g$ the gravitational
constant. In view of the relatively low aspect ratio considered ($A<4$), the Rayleigh
number is based on the height $H$ of the tank, as is most common \citep[see
e.g.][]{Bejan.2013}, allowing also for comparison with other results in the literature.
In this study we consider the linear instability of the steady circulating flow that is
reached at intermediate critical Rayleigh numbers O($10^8$) beyond the onset of the
convective instability for a range of Prandtl numbers. This steady circulating flow is
called later the base flow.

Applications vary with Prandtl number. Generally, higher Prandtl numbers apply to
geophysical flows with Prandtl numbers of 0.7 and 7 for air and water at $20^o\;C$,
respectively, and very high Prandtl numbers for the Earth mantle, with magma
viscosities somewhere around $10^{19}$ \citep[e.g.][]{Busse2006}. For seawater the
Prandtl number ranges from $1<Pr<14$ as a function of temperature and salinity. The
lower Prandtl numbers apply to gases and liquid metals. Atmospheric air has a Prandtl
number in the range of $0.7<Pr<0.79$, Methane gas  in the range of $0.7<Pr<0.87$,
whereas mixtures of liquid Helium may have a Prandtl number between $0.2<Pr<0.6$
depending on its mixtures.  Other applications are semiconductor crystals with $Pr$
O($10^{-2}$) \citep[see e.g.][]{Gelfgat1999}, and nuclear engineering processes that
are associated with convective fluid motions of sodium, lead or alloys for cooling with
$Pr \approx 10^{-1}$ to $10^{-3}$ \citep[see e.g.][]{Grotzbach2013}. Very small Prandtl
numbers apply to astrophysics, stellar and deep solar convection with $Pr \approx
10^{-6}$ \cite[see e.g.][etc.]{Guervilly2019, Pandey2021,Garaud2021}.

In the past, particular attention has been given to the flow transition to a permanent
oscillatory state that occurs in the corner regions of a cavity with $A=1$ for $Pr=0.7$
and a Rayleigh number just above critical, i.e. $Ra\approx 10^8$. The thermal boundary
layers detach and the presence of standing and dissipative internal wave modes were
observed in the interior \citep[see][and references therein]{Paolucci1989, Henkes1990,
LeQuere1998}. Above the critical Rayleigh number, a shear instability occurs in the
vertical boundary layers, with a transition to chaos through quasi periodicity
\citep[see e.g.][]{Lappa2009}. For larger Prandtl numbers differences in behaviour
occur since the thermal boundary layer is thinner with a larger velocity gradient
normal to the boundary, favouring shear instability. Thus, an immediate transition to
turbulence has been observed for Prandtl numbers, $2.5 \leq Pr \leq 7.0$, and a
transition from steady to a periodic state of the jet-like structure for the lower
Prandtl number range $0.25 \leq Pr \leq 2.0$ \citep[see][]{Janssen1995, Chenoweth1986}.

For tall cavities with $1.0 \leq A \leq 3.0$ ($Pr=0.7$) \citep[see e.g.][and references
therein]{Xin2006}, the instability is determined by the detachment of the boundary
layer in the corner regions and the spatial structure of normal modes that fill the
cavity. The instability in the boundary remains relatively small. The inclined flow
structures in the interior that were ascribed to internal waves
\citep[see][]{LeQuere1998,Xin1995} are shown to be in fact part of the unstable mode
\citep{Xin2006}. For cavities with $ A \geq 3$, the traveling waves in the vertical
boundaries have about 10 times higher frequencies with the instability in the vertical
boundary layers being dominant. These wall mode waves occur as Tollmien-Schlichting
waves in the boundary layer for small $Ra$, \citep{Yahata1999,Xin2006,Xin2012}. For
smaller values of $A$, internal waves in the interior dominate the instability
\citep{Yahata1999}. The instability mode is found to be either centrosymmetric or
anti-centrosymmetric, respectively. This instability is part of two Hopf bifurcations
that is encountered for increasing $Ra$ number (and fixed $Pr$ number), with
consecutively the (anti-centrosymmetric or centrosymmetric) internal wave modes, and
for larger Rayleigh number the (anti-centrosymmetric or centrosymmetric) wall modes
\citep[see][]{Burroughs2004,Oteski2015}.

For larger $Ra$, the flow becomes nonlinear with vortices detaching from the boundary
layer and penetrating into the stratified core. These penetrating vortices excite
internal waves with a frequency smaller than the Brunt-Väisäla frequency thus
perturbing the core fluid \citep{Xin1995}. The instability in a rectangular cavity can
thus form in the corner region or in the lateral boundary, and internal waves take part
in the instability. Next to the instability, for certain parameters, a large-scale
circulation has been also observed with the hot boundary layer motion 'connecting' to
the start of the cold boundary layer motion. In the case of conducting horizontal
boundaries this gives rise to limit cycles \citep[see][]{Henkes1990}. A comprehensive
review of studies on vertical convection is given in a historical perspective by
\cite{LeQuere2022}.

Though a range of Prandtl numbers and different aspect ratio have been considered in
the past,  the critical Rayleigh number and the shape and symmetry of the most unstable
corresponding mode are known only for some specific values of $Pr$ and $A$. There are
no clear indications as to the mechanism responsible for the instability, the presence
of locally increasing modes or global modes and their symmetry. In this research, in
order to obtain information about the shape of the most unstable mode, and  to the
critical Rayleigh number, a numerical linear stability analysis is used. The
representation of the perturbations of amplitude, phase and vorticity of each specific
unstable mode allows investigation of the mechanism of instability and the role of
internal waves permitting different flow regimes to be identified.

Apart from the reduced computational costs, the advantage of a two-dimensional approach
is that it is well-posed with a simple geometry and forcing, similar to other flows for
which the knowledge and understanding of regimes and the transition between them is of
fundamental interest. Some well known other examples are Taylor-Couette flow or
Rayleigh-Benard in thin gaps. More related flows are the shear-driven cavity flow (see
e.g. for the homogeneous case \cite{Bengana2019}, and for the stratified case
\cite{Wu2018}), or the self organised state of two-dimensional turbulence on a
rectangular domain interacting with vorticity generated at the slip-free boundaries
\citep{Konijnenberg1998, vanHeijst2006}. In the latter example, a large central vortex
interacts with the boundaries, implying aspects of symmetry with flow phenomena
analogous to large cell flows in vertical convection, but without baroclinic effects.

In a three dimensional box with periodicity in the third direction \citep[see][for
$Pr=0.7$]{Xin2012}, the instability starts for a 10 times smaller Rayleigh number. The
three dimensional effects are however modest, with low frequency modes loosing their
stability earlier than in the two-dimensional case. Two dimensional simulations are
found to be also satisfactory for larger aspect ratio ($A=4$) and capture the general
features of buoyancy-driven flow, as long as it is not turbulent, i.e. up to $Ra \sim
10^{10}$ \citep{Trias2007}. Also partial similarities with the two-dimensional
counterpart have been noticed in cubic cavities \citep[see][]{Gelfgat2017,
Gelfgat2020a, Gelfgat2020b}. This is further discussed at the end of the conclusions.

In the next section, \S \ref{sec:num}, the numerical code with the linear approach and
the decomposition into leading modes are discussed next to the nonlinear approach and
the diagnostics. In the subsequent section, \S \ref{sec:results}, the results are
presented with the different base states, leading linear modes, and the different
observed regimes. In \S \ref{sec:conclusions}, the main conclusions are presented and
further discussed.



\begin{figure}
\centerline{
\includegraphics[width=0.48\textwidth]{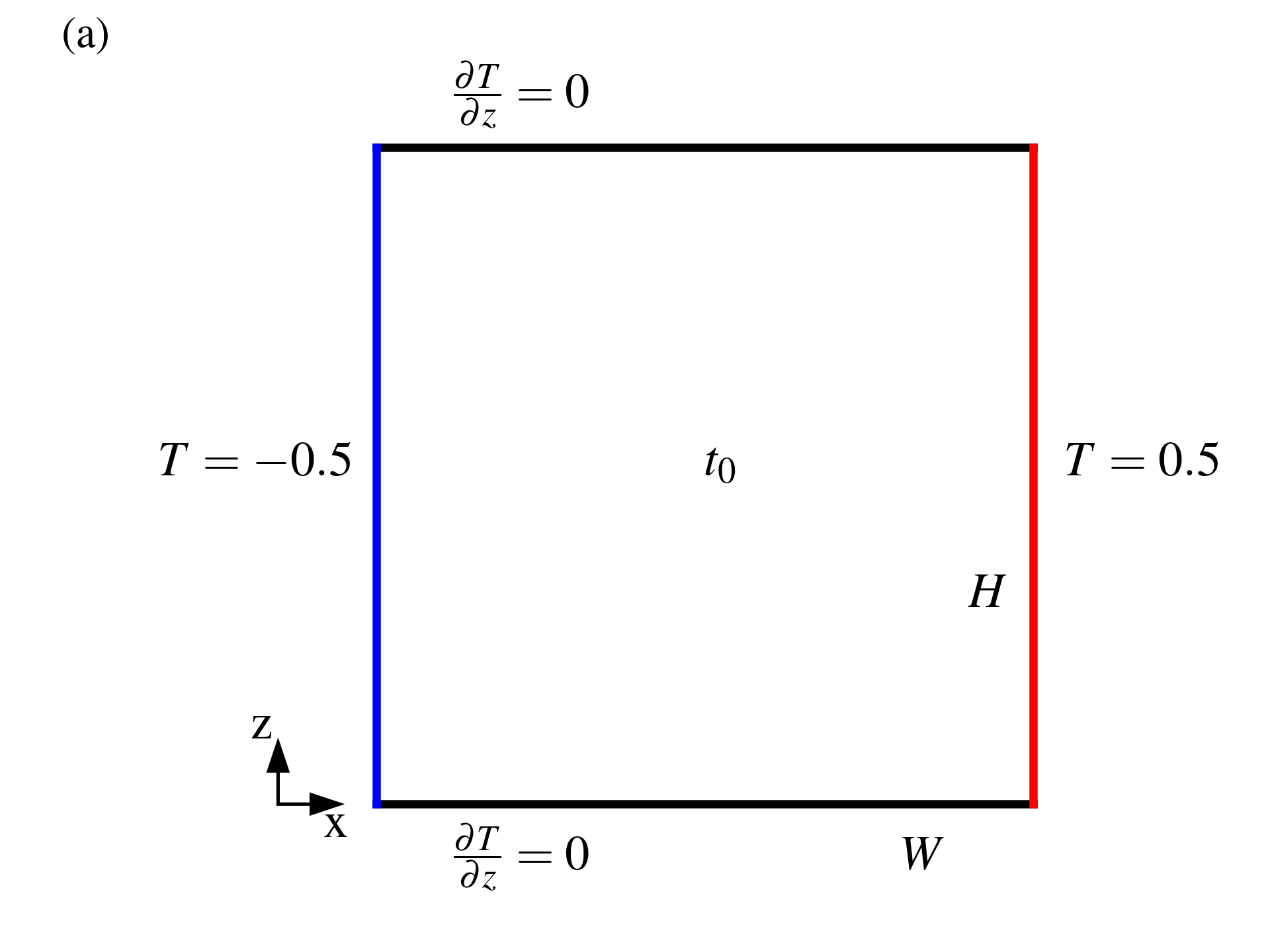}
\includegraphics[width=0.48\textwidth]{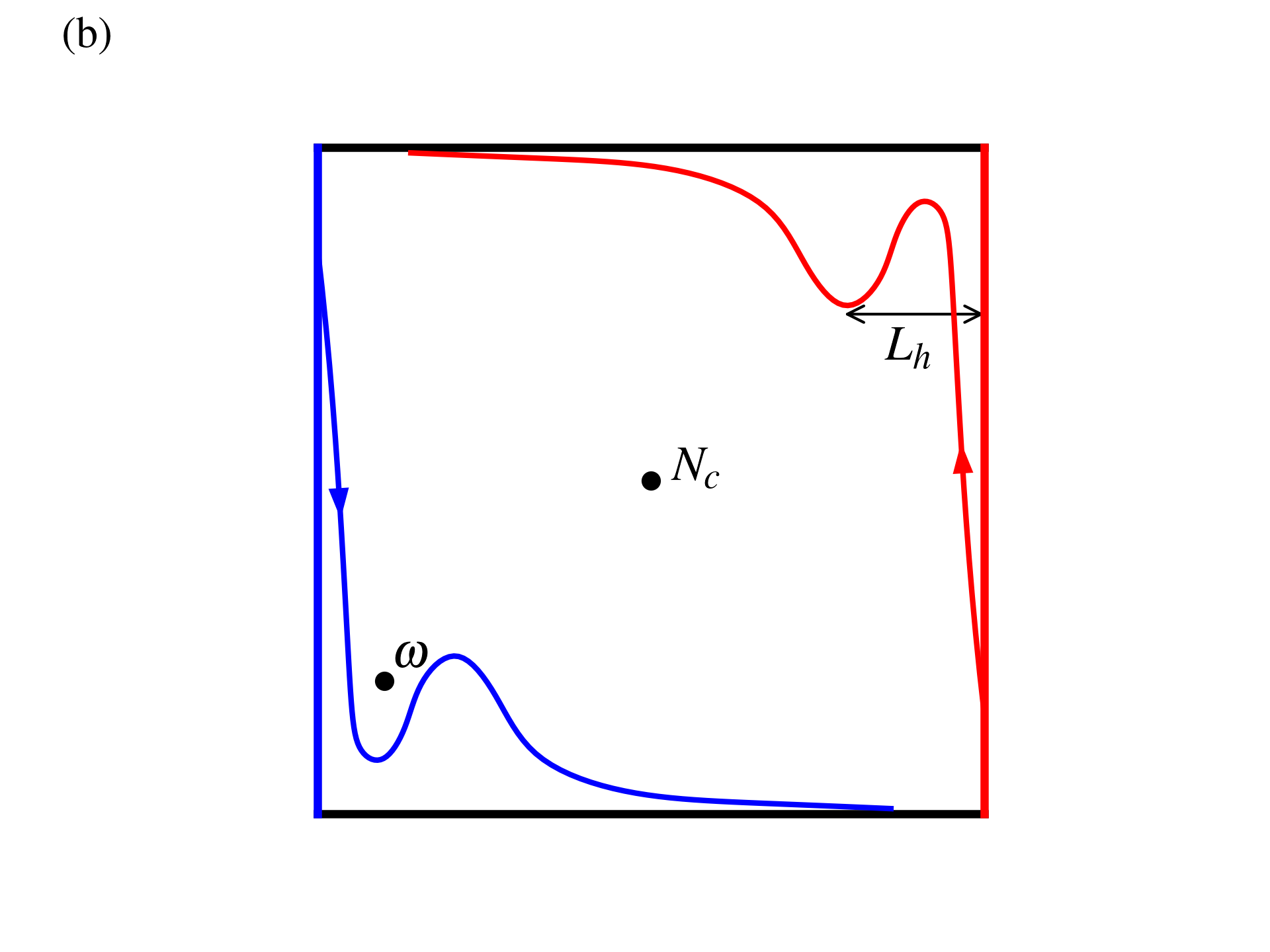}}
\caption{(a) Schematic of the problem with the walls kept at constant temperatures 0.5
and -0.5 so that $\Delta T = 1.0$. The initial condition for the nonlinear simulations
is $T_0= \Delta T \times \left ( x/x_{max} - 0.5 \right ) + noise$ and $(u, v) =
(0,0)$, and for the linear simulations, $T_{p_0}= noise$ and $(u_p, v_p) = (0,0)$. (b)
Location of the interrogation points used to assess the state of the flow with the
oscillation frequency $\omega$ measured in the corner at $(x,z)=(0.1,0.1)$, for $A=1$,
$H=$(0,1) and $W=$(1,0), $N_c$ the stratification in the centre of the cavity. The
maximum speed in the buoyancy current is sketched by red and blue curved lines. $L_h$
is the horizontal distance from the wall to the minimum (in $z$) on the red line of the
current at the top. \label{fig:sketch}}
\end{figure}
\section{Numerical setup}
\label{sec:num}

\subsection{Governing equations and linear stability approach}

We consider a two-dimensional flow inside a rectangular cavity of aspect ratio
$A_v=H/W$ with cavity height $H$ and width $W$, adiabatic top and bottom boundaries,
and the two walls kept at a constant temperature with temperature difference $\Delta T$
(see figure \ref{fig:sketch}(a)). The scales of this problem are $ \Delta T$ for the
temperature difference and for the length scales $(x,y) \sim (\delta_{T},H)$ where
$\delta_{T} = (\kappa t_r)^{1/2}$ is the thickness of the heated boundary layer, and
$t_r$ is the reference time defined below. Using the momentum equations, the friction
term then scales with the buoyancy term, i.e, $\nu V_r/(\delta_{T})^{2} \sim g \beta
\Delta T$ which yields a characteristic speed $V_{r} = \left(\kappa / H \right)
Ra^{1/2} $. The time scale $t_{r} = H/V_r$ is obtained from the balance between the
advective term and diffusive term in the temperature equation. With these scales one
obtains for the dimensionless form of the continuity equation, Boussinesq approximation
of the Navier-Stokes equations, and the temperature equations, respectively,

\begin{equation}
\mathbf{\bnabla} \cdot \mathbf{v} = 0,
\label{eq:pertemp}
\end{equation}

\begin{equation}
\frac{\partial \mathbf{v}}{\partial t} + \mathbf{v} \cdot \mathbf{\nabla \mathbf{v}} = -\mathbf{\nabla}P + \frac{Pr}{Ra^{1/2}} \mathbf{\nabla}^{2}\mathbf{v} + Pr \Theta \mathbf{e}_{z},
\label{eq:pertu}
\end{equation}

\begin{equation}
\frac{\partial \Theta }{\partial t} + \mathbf{v} \cdot \mathbf{\nabla}\Theta = \frac{1}{Ra^{1/2}}\mathbf{\nabla}^{2} \Theta,
\label{eq:pertw}
\end{equation}

\begin{figure}
\centerline{
\includegraphics[width=0.6\textwidth]{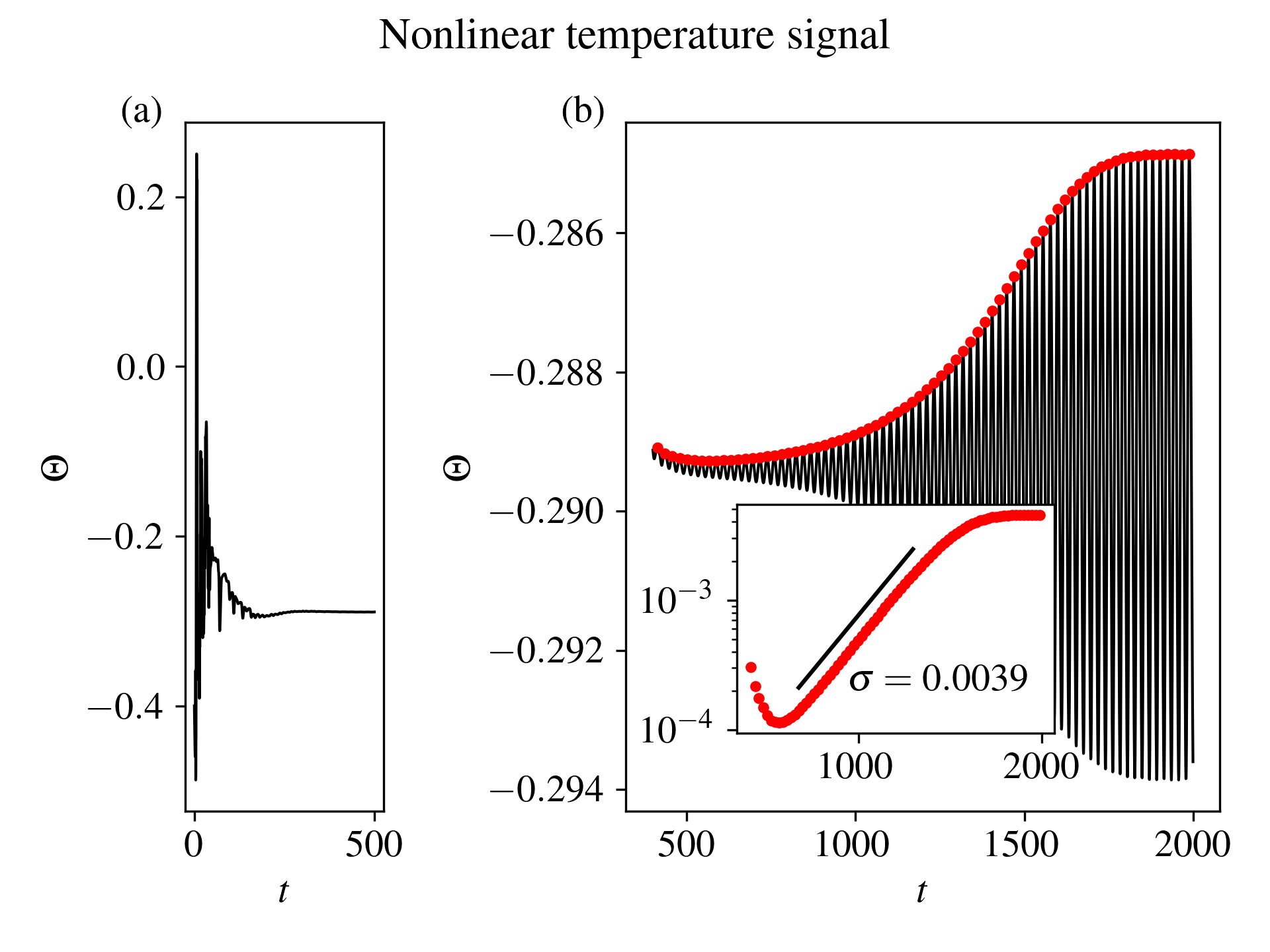}
}
\caption{Temperature signal of a nonlinear simulation at point (x, z) = (0.1, 0.1) for
$A=1.0$, $Pr=0.71$ and $Ra = 1.85 \times 10^{8}$, with (a) the evolution of the flow
from $t=0$, and (b) the evolution to a steady state and subsequent exponential growth
in amplitude and saturation. The inset shows the perturbation in log scale with the fit
giving the growth rate (straight line).} \label{fig:NLsignal}
\end{figure}

\noindent with $\mathbf{v}$ the velocity vector, pressure $P$, and $\Theta$ the
dimensionless temperature $\left(T - T_{r} \right)/ \Delta T $ with $T_r$ being the
average temperature (here $T_r=0$). As mentioned above, the control parameters of this
flow are the Rayleigh number, Prandtl number and aspect ratio $A=H/W$ which depends on
the chosen length $W$ since the heigh $H$ is kept constant.

The no-slip condition is used for all boundaries. For the temperature, the Dirichlet
condition is used with $\Theta=0.5$ and $\Theta=-0.5$ on the two lateral walls
respectively, and for zero heat flux the Neumann condition at the two horizontal
boundaries.

Numerical simulations are performed with the spectral element code
Nek5000\footnote{\url{https://nek5000.mcs.anl.gov}} using the two newly developed
Python packages Snek5000\footnote{\url{https://snek5000.readthedocs.io}} and
Snek5000-cbox\footnote{\url{https://github.com/snek5000/snek5000-cbox}}
\cite[]{mohanan_snek5000_2023, fluiddyn, fluidsim}. The packages are available online
(see footnotes) and the data that we have produced here is available as a Zenodo
dataset\footnote{\url{https://zenodo.org/record/7827872}}.

Tests have been conducted for $Pr=0.71$ and $A=1$ and results for the growth rate and
oscillation frequency have been validated with respect to former studies in the
literature.  The resolution employed (see table I Appendix A) allows for the study of
the motion in the thin boundary layers near the walls, and accuracy in growth rate and
oscillation frequency.
Figure \ref{fig:NLsignal} shows a typical temperature signal measured in the corner of
the cavity (see figure \ref{fig:sketch}(b)) for the nonlinear simulation. The large
difference in temperature between figures \ref{fig:NLsignal}(a) and (b) reveals the
transient to the base state ($t < 250$) and the subsequent growth and saturation of the
instability ($t>250$), respectively. From $t$=0 to 250, the motions along the
boundaries and the stratification in the interior develop. The base state is a steady
state with minimum amplitude of oscillations (at approximately $t \approx 500$), with
motions in thermal boundary layers in the presence of a stratified interior. From $t =
500$ onwards, a linear instability leads to the exponential growth of the amplitude of
the oscillation up till about $t=1375$, after which it saturates and small nonlinear
oscillations in amplitude develop. The oscillation frequency shows a perfect
exponential growth in the range of about $500<t<1400$ (see the inset in figure
\ref{fig:NLsignal}(b)).

For each set of control parameters $Pr$ and $A$, nonlinear simulations are performed to
obtain a first approximation of the critical Rayleigh number $Ra_c$. Nonlinear
simulations reach a steady state at Rayleigh number of O($\sim 10^8$) but still smaller
than the critical Rayleigh number, i.e. $Ra<Ra_c$, while for slightly higher $Ra$
numbers, the flow starts to oscillate at $t \approx 500$. Three nonlinear simulations
are performed for three values of $Ra$ slightly larger than the estimated $Ra_c$ using
the Selective Frequency Damping (SFD) method of \cite{Aakervik2006}, which give us
three steady base states.  This method has also been used recently for computing the
base state of the flow that is induced when tilting a cavity containing a stably
stratified fluid \citep[see][]{Grayer2020}.

Subsequently, the linear stability of these steady base flows is considered. Using
perturbed variables $\Theta = \Theta_{b} + {\theta}'$, $\mathbf{V} = \mathbf{V}_{b} +
{\mathbf{v}}'$, $P = P_{b} + {p}'$ where subscript $b$ represents the base state and
superscript ${}'$ the perturbation, and neglecting second order terms, we obtain
linearised perturbation equations of the form

\begin{equation}
\frac{\partial{\mathbf{v}}'}{\partial t} + \mathbf{V_{b}} \cdot \mathbf{\nabla{\mathbf{v}}'} + {\mathbf{v}}' \cdot \mathbf{\nabla \mathbf{V_{b}}} =
 -\mathbf{\nabla}{p}' + \frac{Pr}{Ra^{1/2}} \mathbf{\nabla}^{2}{\mathbf{v}}' + Pr {\theta}' \mathbf{e}_{z},
\end{equation}

\begin{equation}
\frac{\partial{\theta}' }{\partial t} + \mathbf{V_{b}} \cdot \mathbf{\nabla}{\theta}' + {\mathbf{v}}' \cdot \mathbf{\nabla}\Theta_{b} =
 \frac{1}{Ra^{1/2}}\mathbf{\nabla}^{2} {\theta}'.
\end{equation}

Linear simulations are run for the three steady base states obtained from the SFD
method, and corresponding to three unstable $Ra$ values. A small amount of noise of the
order of $10^{-6}$ is added so that, due to the linear instability, an exponential
growth of the leading mode is observed in the whole cavity. This noise is random and
therefore not necessarily symmetric.  Then, the growth rate is determined for each
linear simulation, thus providing eventually three different growth rates. Using a
linear interpolation of these growth rates,  the critical Rayleigh number $Ra_c$ is
extrapolated from the value for zero growth-rate. The base flow and the perturbation
analysed in the next section are obtained from the nonlinear and linear simulations for
the Rayleigh number just above this critical Rayleigh number.

\subsection{Decomposition of the leading linear mode}
\noindent In view of former observations of this instability \citep[see
e.g.][]{Xin2006}, the different variables can be decomposed during the oscillating
exponential growth into
\begin{equation}
\theta'\left(x,z,t\right)= A_{\theta}(x,z) \cos\left(\omega t + \Phi_{\theta}\left(x,z\right)\right)e^{\sigma_{r}t},
\end{equation}

\begin{equation}
u'\left(x,z,t\right)= A_{u}\left(x,z\right)\cos\left(\omega t + \Phi_{u}\left(x,z\right)\right)e^{\sigma_{r}t},
\end{equation}
\noindent and,
\begin{equation}
w'\left(x,z,t\right)= A_{w}\left(x,z\right)\cos\left(\omega t + \Phi_{w}\left(x,z\right)\right)e^{\sigma_{r}t},
\end{equation}

\noindent where $A(x,z)$, $\omega$, $\Phi(x,z)$, and $\sigma_{r}$ are amplitude,
frequency, phase, and growth rate of the field variables, respectively. The growth rate
$\sigma$ and the oscillation frequency $\omega$ are computed by an algorithm based on
Hilbert transforms. The amplitude fields are then obtained by taking the time maximum
of the perturbation variables divided by $e^{\sigma_{r}t}$. Finally, the phase fields
are obtained with one curve fit per grid point and variable.

\section{Numerical results}
\label{sec:results}

\subsection{The steady base flow and diagnostics}
\label{sec:diagnostics}

\begin{figure}
\centerline{
\includegraphics[width=\textwidth]{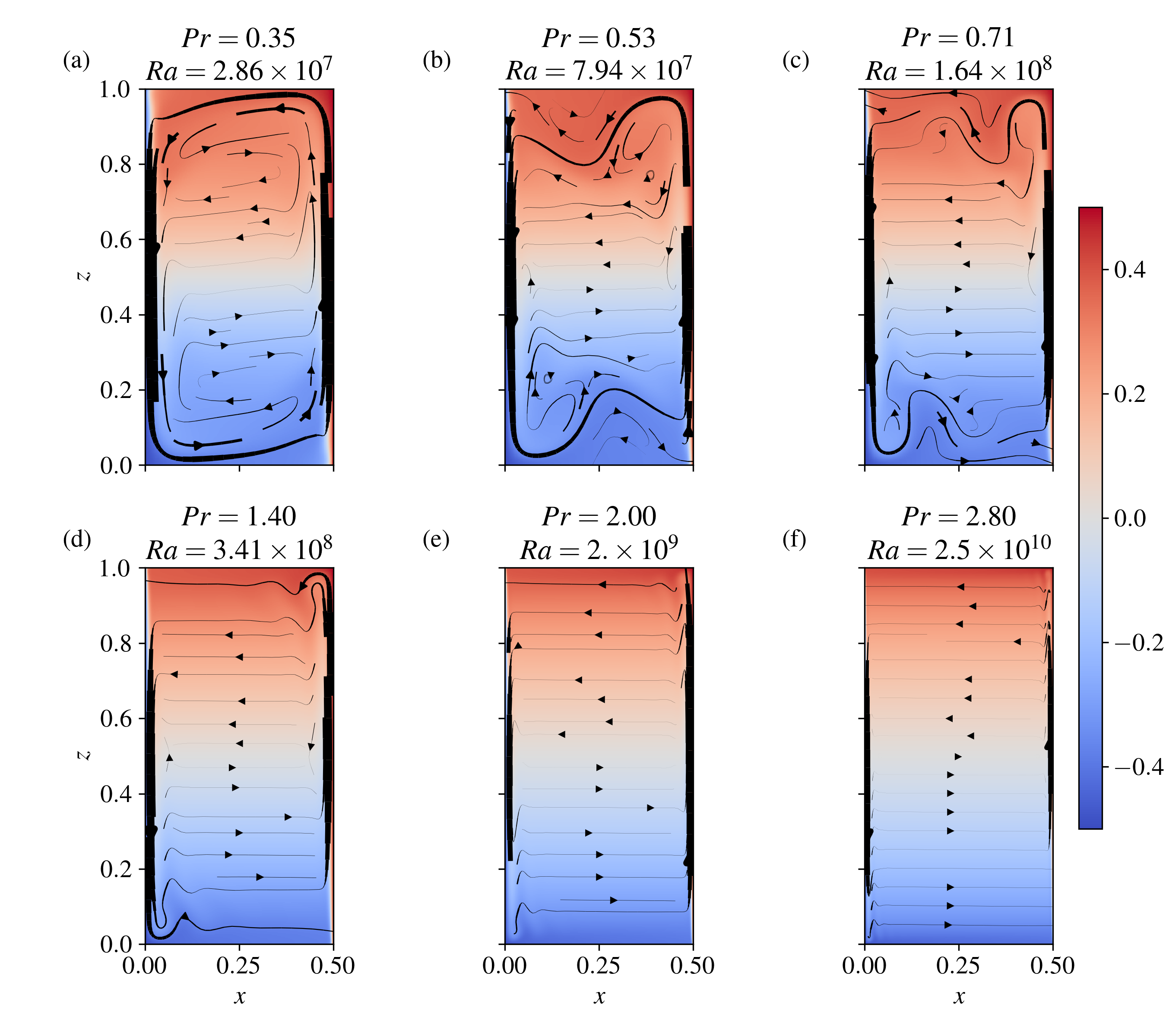}
}
\caption{The base flow states for $A=2.0$ and different $Pr$: (a) $Pr=0.35$, (b)
$Pr=0.53$, (c) $Pr=0.71$, (d) $Pr=1.4$, (e) $Pr=2.0$ and (f) $Pr=2.8$. Stream lines are
supplied with arrows indicating flow direction, and color represents the temperature.}
\label{fig:base_state_stream_A_2.0}
\end{figure}

\begin{figure}
\centerline{
\includegraphics[width=0.8\textwidth]{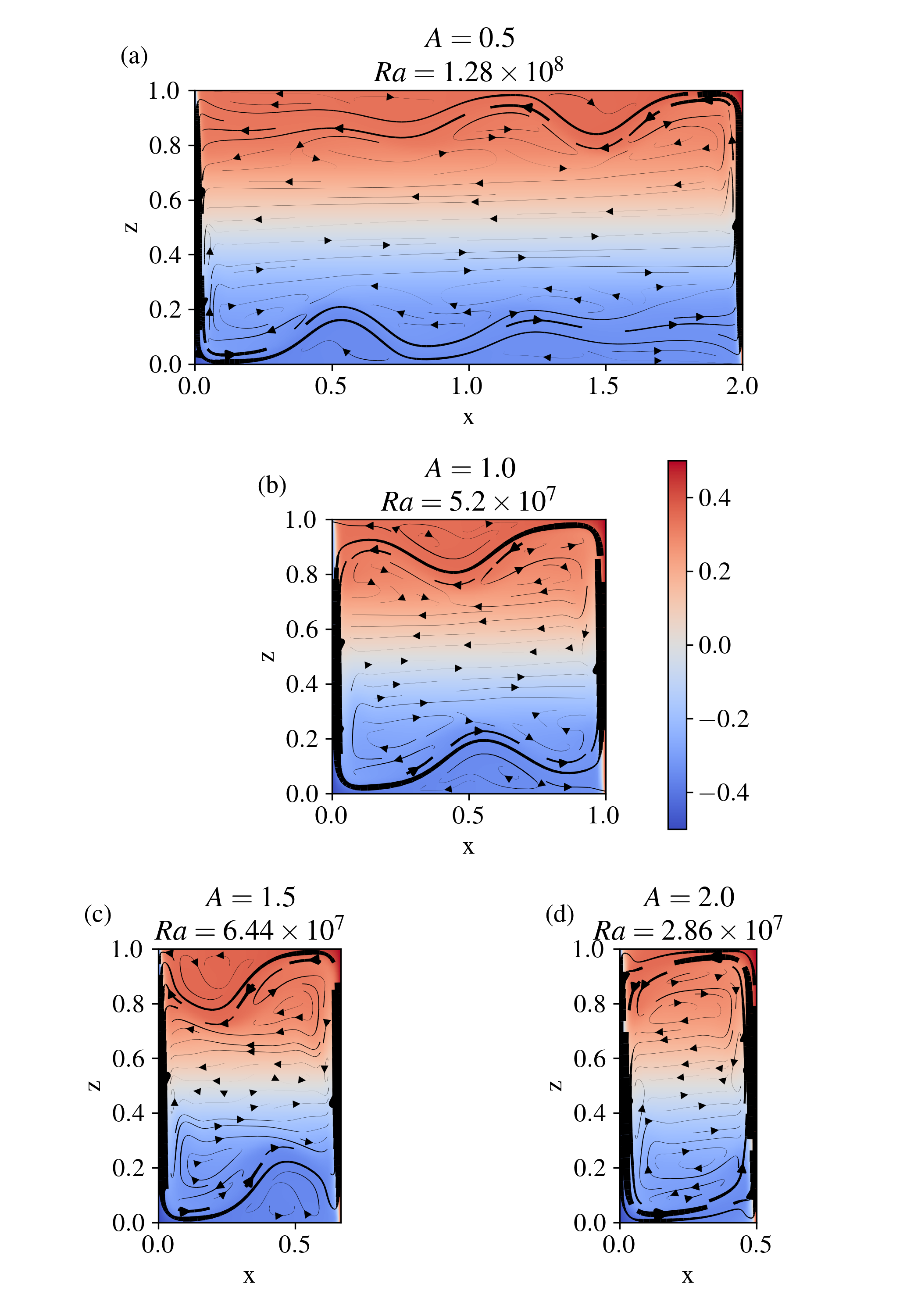}
}
\caption{The base flow state for $Pr=0.35$ and different $A$: (a) $A=0.5$, (b) $A=1.0$,
(c) $A=1.5$ and (d) $A=2.0$. Lines show the stream lines with the arrows the flow
direction, and color represents the temperature.} \label{fig:base_state_stream_Pr0.35}
\end{figure}

Figure \ref{fig:base_state_stream_A_2.0}(a-f) shows the steady base flow for different
Prandtl numbers and constant aspect ratio $A=2.0$. From figures
\ref{fig:base_state_stream_A_2.0}(a-f) one notices that for larger Prandtl numbers the
buoyancy currents detach and meander along the horizontal boundaries, the number of
meanders depending on the $Pr$ number. Clear experimental and numerical support for
this meandering is shown by \cite{Xu2008}. The wave length of this meandering buoyancy
current changes significantly with $Pr$, while its presence is limited by the
horizontal extend $W$ of the cavity represented by $A$ (note $H$ is kept constant).
\begin{figure}
\centerline{
\includegraphics[width=0.48\textwidth]{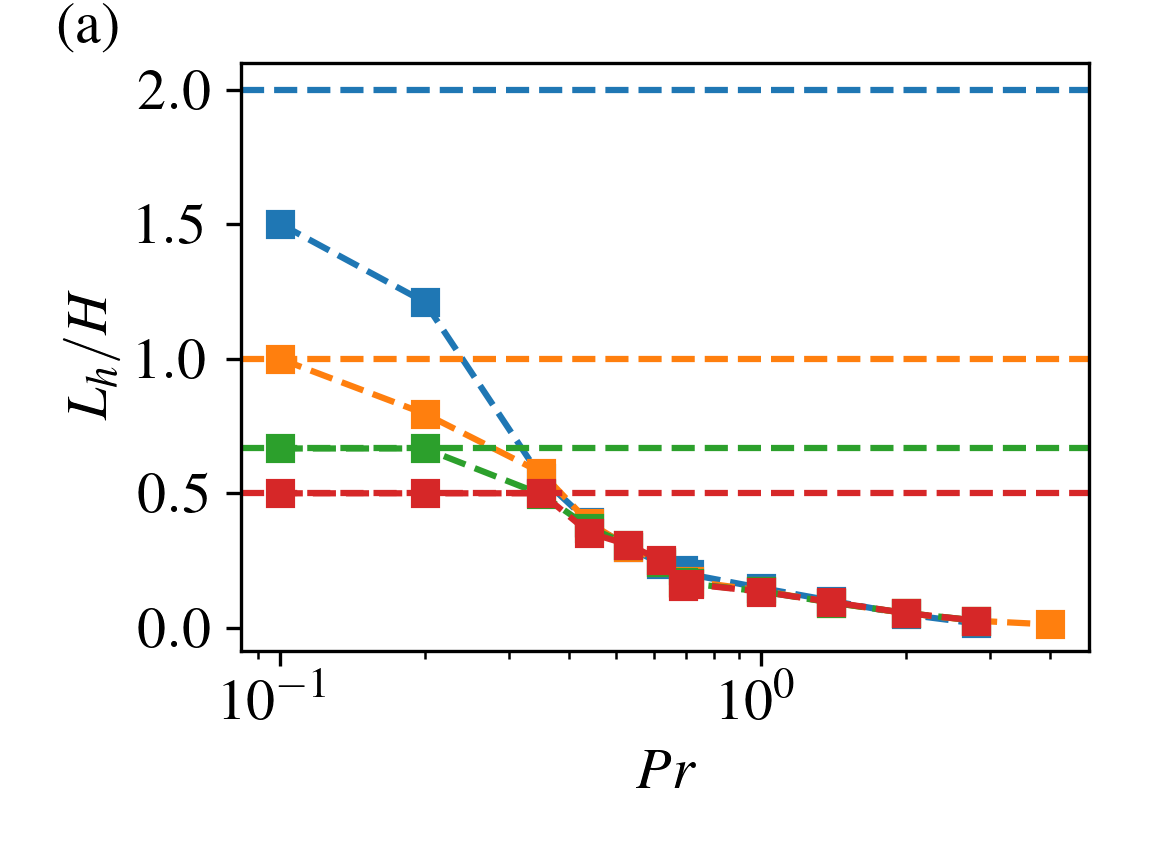}
\includegraphics[width=0.48\textwidth]{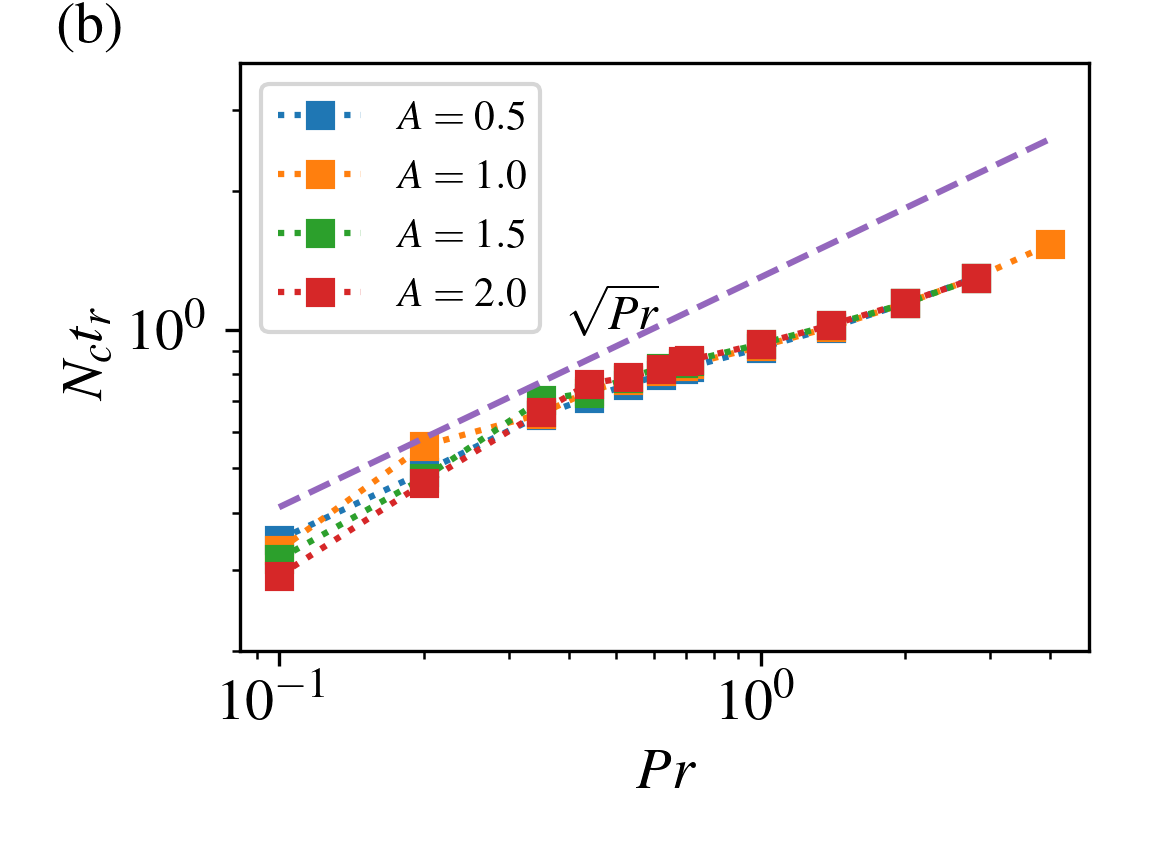}}
\caption{(a) Meander length scale $L_h/H$ against Prandtl number $Pr$, with the dashed
lines representing the aspect ratios $A=H/W$, and (b) $N_c$ Brunt-V\"ais\"al\"a
frequency scaled with the characteristic time $t_r$ at the center of cavity as a
function of $Pr$ number.  For comparison the slope of $\sqrt{Pr}$ predicted by
\ref{eq:strat} is shown by the dashed line.  } \label{fig:Lh_vs_Pr}
\end{figure}
\noindent When the width of the cavity is smaller than this wave length, the head of
the buoyancy current joins the start of the cold boundary layer, and vice versa near
the bottom boundary, such that the two currents reinforce each others inertia leading
to a large scale, fast circulation along the boundaries (see figure
\ref{fig:base_state_stream_A_2.0}(a)). With increasing $Pr$ number, ($Pr<0.7$ in figure
\ref{fig:base_state_stream_A_2.0}), the large-scale circulation decreases in strength
and the buoyancy current detaches from the horizontal boundary, i.e. $L_h<W$, and it
meanders locally. For larger $Pr$ numbers, ($Pr>1$, see figures
\ref{fig:base_state_stream_A_2.0}(d,e,f)) a horizontal exchange flow establishes in the
interior between the two thermal boundary layers, The boundary layers are thinner for
these higher $Pr$ numbers. For a constant $Pr$ it depends on the aspect ratio whether
there are cell patterns (as in figure \ref{fig:base_state_stream_A_2.0}(a)) or rather
horizontal exchanges between the thermal boundary layers (as in figure
\ref{fig:base_state_stream_A_2.0}(f))  \citep[see][]{Xin2006}.

Figure \ref{fig:base_state_stream_Pr0.35} shows flows for $Pr=0.35$ and varying aspect
ratio $A$. When the width of the cavity is large (see figure
\ref{fig:base_state_stream_Pr0.35}(a) for $A=0.5$), the buoyancy current meanders along
the horizontal boundary. With increasing aspect ratio $A$, the meander length-scale of
the buoyancy current becomes smaller than the width of the cavity, and there is an
increasing tendency for the formation of cell circulation. Two circulation cells appear
for $A=2$ in figure \ref{fig:base_state_stream_Pr0.35}(d).

In all cases, a stable density gradient $ -\beta d T/d z$ is present in the interior.
The buoyancy (Brunt-V\"ais\"ail\"a) frequency scaled with time $t_r = H^2 /
(\kappa{Ra}^{1/2})$ is given by
\begin{equation}
(N_c t_r)^2 = g \beta \frac{d T}{dz} \frac{H^4}{\kappa^2}
\frac{1}{Ra} = Pr \frac{H}{\Delta T} \frac{dT}{dz},
\label{eq:strat}
\end{equation}
showing that the Prandtl number, height of the reservoir and density gradient are
relevant for the scaled buoyancy frequency, with slightly stronger stratifications for
shallow cavities  (small $A$) and weaker stratifications for high and narrow cavities
(large $A$).

In the simulations, the buoyancy frequency $N_c$ is determined by the density gradient
in a small region around the centre of the tank.  The meander length-scale of the
buoyancy current, $L_h$, is defined as the horizontal distance from the wall where the
line that describes the maximum speed (see Figure \ref{fig:sketch}(b)) has a minimum in
$z$. $L_h$ has been determined from the flow near the top.  In the perturbed state, the
instability causes oscillations in the temperature with frequency, $\omega$. Its
amplitude increases due to the linear instability, and though present in the entire
tank, they are mainly visible in the corner regions where the thermal motion along the
wall is blocked by the horizontal boundary \citep[see e.g.][and references
therein]{LeQuere1998, Xin2006}. Thus, to analyse this flow, the frequency of the
oscillation frequency of the instability mode, $\omega$, the density stratification in
the interior, $N_c$, and the wave length of the meandering buoyancy current, $L_h$, are
measured at the locations shown in Figure \ref{fig:sketch}(b).

When the temperature oscillations have a higher frequency than the buoyancy frequency,
i.e. $\omega>N_c$, internal waves cannot propagate in the interior and are evanescent.
In contrast, when $\omega<N_c$, the buoyancy currents near top and bottom boundaries
may couple due to the internal waves that propagate into the interior. Even though a
larger value of $N$ may exist near the top and bottom boundaries, the value $N_c$ in
the centre of the tank is taken as reference value since it is more relevant for the
coupling of top and bottom regions. As mentioned, flows with $\omega/N_c>1$ are called
fast, and flows with $\omega/N_c<1$ slow since allowing for the propagation of internal
gravity waves, and, in most cases, for the coupling between the two buoyancy currents.
In the absence of internal waves, top and bottom instability generally start to grow
independently with different perturbation amplitude, resulting in asymmetry. This is
referred to as amplitude asymmetry, or shortly asymmetry and is further detailed below.

For the base states, figure \ref{fig:Lh_vs_Pr}(a) shows the (scaled) wave length of the
meandering buoyancy current $L_h/H$ against $Pr$ for different aspect-ratio of the
cavity. For small $Pr$ there is no detachment and $L_h$ is larger than the width of the
cavity, i.e. $L_h=1/A$. In this case, the horizontal buoyancy currents and boundary
layers at the side walls reinforce each other leading to a large cell circulation. For
larger $Pr$ numbers, the boundary layers are thin and the aspect ratio $A$ has no
influence on the value of $L_h/H$. The effects obtained for decreasing $Pr$ can also be
obtained for larger aspect ratio. Boundary layers grow with distance and become thicker
and have more inertia, leading to relatively larger values of $L_h$.

Figure \ref{fig:Lh_vs_Pr}(b) shows that the stratification in the interior increases
less for $Pr>0.4$ than for $Pr<0.4$ since larger gradients form near the horizontal
boundaries, and a weaker stratification forms in the interior. The aspect ratio $A$ has
an influence on the internal stratification only for smaller Prandtl numbers for which
the large cell circulation dominates.

The asymmetry in the value of the amplitude between the top and bottom currents is
considered in particular since its relation with the presence of internal waves is
novel and provides new insights. For both currents the growth rates are the same, but
due to an asymmetry in the initial noise the amplitudes of the perturbations may be
different. Writing out the equations for the temperature, we have then,
\begin{equation*}
\theta' (\mathbf{x},t) = \left( C_+A_+(\mathbf{x}) +C_- A_-(\mathbf{x}) \right)
\cos(\omega t + \phi (\mathbf{x})) \,e^{\sigma t},
\end{equation*}
with the ratio $C_+/C_-$ depending on the initial noise. Thus, in contrast to the
coupled case with top and bottom currents having the same amplitude and internal waves
being part of the same global mode, we have in the uncoupled case two local regions
that are growing independently.

The symmetry conditions are imposed by the boundary conditions, so that for a solution
$\Phi$ with
\begin{equation*}
\Phi(\mathbf{x})=
\begin{pmatrix}
\Theta_b(\mathbf{x})\\ \mathbf{V_b}(\mathbf{x}) \\ P_b(\mathbf{x})
\\ \xi_b(\mathbf{x})
\end{pmatrix}
\end{equation*}
with $\xi_b=\nabla \times \mathbf{V_b}$ the vorticity in the basic state, the equations
are invariant for the transformations
\begin{equation*}
R\,\Phi(\mathbf{x})=
\begin{pmatrix}
-\Theta_b(\mathbf{-x})\\ -\mathbf{V_b}(\mathbf{-x}) \\ P_b(\mathbf{-x}) \\
\xi_b(\mathbf{-x}).
\end{pmatrix}
\end{equation*}
In case there is symmetry, the solution can be either "centro-symmetric", i.e.
$R\,\Phi(\mathbf{x})=\Phi(\mathbf{x})$, or "anti centro-symmetric", i.e.
$R\,\Phi(\mathbf{x})=-\Phi(\mathbf{x})$ \citep[see][]{Burroughs2004}. Counter
intuitively, for centro-symmetry, the velocity phase must be opposite to that in the
other corner whereas the vorticity must be equal, and vice versa for anti
centro-symmetry.  This centro-symmetry was tested considering the point reflected value
with respect to the centre of the tank. Thus, the nondimensional temperature in the top
half of the cavity $\theta'_{\text{top}}$ is compared with its flipped counterpart in
the bottom half of the cavity $\theta_{\text{bot}}^{'\text{flip}}$. The parameters for
centro -symmetry and anti centro-symmetry then become, respectively,
\begin{equation*}
I_C = \frac{\langle \left| \theta'_{\text{top}} - \theta_{\text{bot}}^{'\text{flip}} \right| \rangle}{\sqrt{\langle\theta'^2\rangle}},
\end{equation*}
\begin{equation*}
I_A = \frac{\langle \left| \theta'_{\text{top}} + \theta_{\text{bot}}^{'\text{flip}} \right| \rangle}{\sqrt{\langle\theta'^2\rangle}},
\end{equation*}
with the brackets standing for the average over the domain, and the enumerator
representing the rms-value of $\theta'$. When the perturbation is centro-symmetric or
anti-centro-symmetric, either $I_C$ or $I_A$, respectively, is small and never zero
because of numerical noise. Both values are large when there is no such symmetry. To
identify centro-symmetry,   the difference of the reciprocals of $I_C$ and $I_A$
\begin{equation}
I_{dif} = \frac{1}{I_C} - \frac{1}{I_A},
\label{centro}
\end{equation}
is used, with anti centro-symmetry for $I_{dif} > 0$, i.e. the temperature
perturbations in the top and bottom halves of the cavity have the same sign, and
centro-symmetry for $I_{dif} < 0$, i.e. opposite temperature perturbations in the top
and bottom halves of the cavity. When $I_{dif} \simeq 0$, there is no symmetry, i.e.
the top and bottom currents are asymmetric in amplitude. The expression (\ref{centro})
therefore provides simultaneously information about centro symmetry and amplitude
symmetry.

Centro-symmetry is generally better solved with different methods that provide a
continuous distribution for the critical Rayleigh number and the centro-symmetry as a
function of Prandtl number \citep[see e.g.][and references therein]{Lyubimova.2009}.
Since this was not the aim of the present investigation, the centro symmetry below is
provided for completeness, and the details are left for future study.

Below, the six observed regimes for increasing $Pr$ are discussed for a single aspect
ratio $A=1$. For the regimes shown in figures \ref{fig:Group_picture1} the field of the
base flow is shown next to the fields of the perturbations predicted by the linear mode
decomposition, with (b) amplitude, (c) phase, and (d) vorticity. The perturbation of
the vorticity field (d) reveals the spatio-temporal nature of the disturbing wave
packets. The phase map (c) shows the distribution of the length scales in the field and
can be considered as a signature  of the internal waves. In addition, two time steps
show the base flow perturbed with the linear perturbation (see figure
\ref{fig:Group_picture2}(a,b)).

Movies of the unsteady states are provided as supplementary material. The effect of
varying aspect ratio $A$, along with the measured parameters $\omega/N_c$ and the
regime diagram in the space set by $A$ and $Pr$ are discussed below.

\newpage \subsection{Six unsteady regimes} \label{sec:regimes}

\begin{figure}
\centerline{
\includegraphics[width=0.97\textwidth]{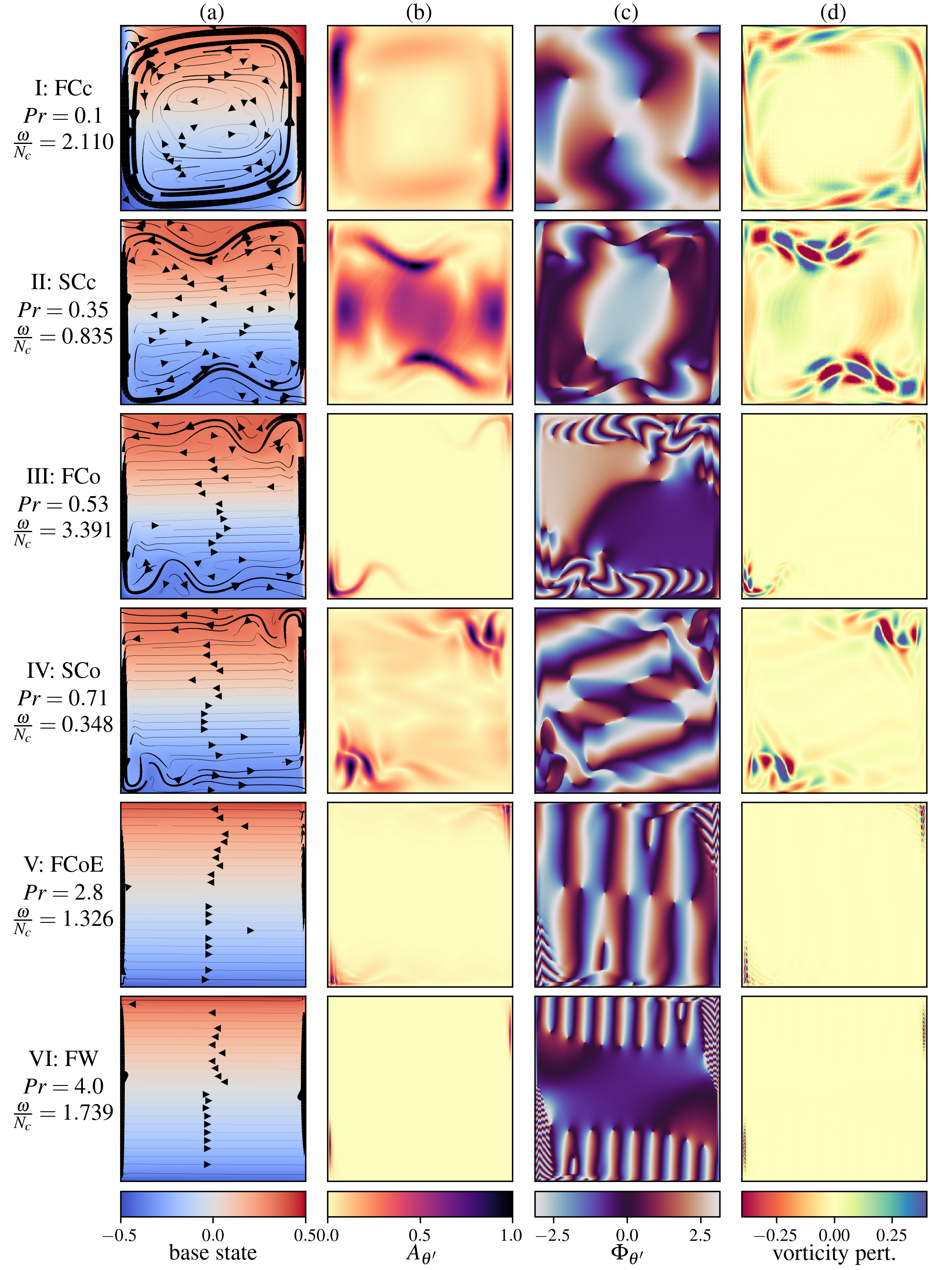}}
\caption{Flow maps for $A=1.0$ and different different values of Prandtl number $Pr$
and oscillation frequency $\omega/N_c$, (case I-VI) with (a) base state (b) temperature
amplitude $A_{\theta’}$ (c) phase map $\Phi_{\theta’}$, and (d) vorticity perturbation
$\xi’$. Supplementary movies of the regimes I Fast Circulation cells (FCc) (Movie 1), II Slow
Circulation cells (SCc) (Movie 2), III Fast Corner flow (FCo) (Movie 3), IV Slow Corner flow (SCo) (Movie 4),V Fast
Corner with Evanescent waves (FCoE) (Movie 5) and VI Fast instabilities at the Wall (FW) (Movie 6) are
available on line. } \label{fig:Group_picture1}
\end{figure}

\begin{figure}
\centerline{
\includegraphics[width=0.75\textwidth]{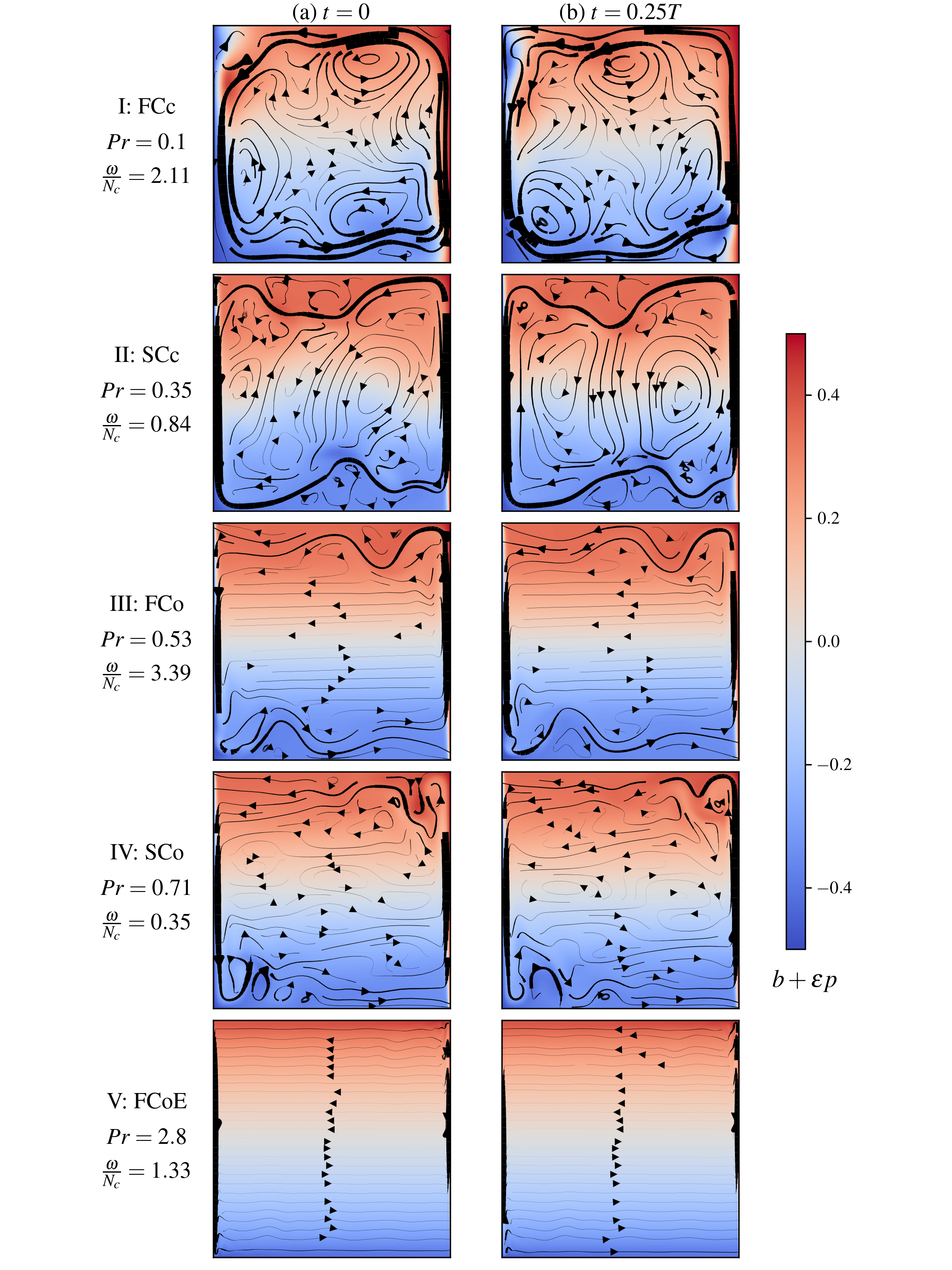}}
\caption{The base flows corresponding to regimes shown in figure
\ref{fig:Group_picture1} plus the perturbation at (a) $t=0$ and (b) $t=0.25T$. Arrows
on streamlines indicate flow direction and color represents temperature. Case VI is not
shown since differences with case V are very small. For all cases I-VI, the same movies as referred to in figures
\ref{fig:Group_picture1}, are represented on line. }
\label{fig:Group_picture2}
\end{figure}

The identification of the different regimes, shown in figure \ref{fig:Group_picture1},
is based on the detachment of the buoyancy current from the top and bottom boundary
and/or the location of the instability, the existence of a large scale circulation, and
whether the scaled oscillation frequency is slow ($\omega/N_c \leq 1$) or fast
($\omega/N_c>1$) allowing for presence of internal waves (or not).  With increasing
Prandtl number, the plumes detach earlier from the horizontal boundaries, and the
presence of internal wave motions in the interior changes accordingly (see figures
\ref{fig:Group_picture1} I-VI); Figures \ref{fig:Group_picture2} I-V show the
transition from a convective regime with circulation cells moving locally through the
tank to flows confined to boundary currents with horizontal exchange between them.
Below we present each flow regime.

In case I ($Pr=0.1$) (see figure \ref{fig:Group_picture1}), there is no detachment of
the buoyancy current from the horizontal boundaries and the hot buoyancy current joins
the cold thermal boundary layer resulting in a large cell circulation \citep[see
e.g.][]{Xin2006}, thus coupling the top and bottom motions. The phase plot (figure
\ref{fig:Group_picture1}(c) I) shows a central rotary motion. In the perturbed flow
(see figures \ref{fig:Group_picture2}(a,b) I) smaller cells move in the interior with
the large cell circulation showing that convective motions dominate. The oscillation
frequency of this mode is larger than the buoyancy frequency, $\omega/N_c=2.26$, and
internal waves can therefore not propagate. This flow regime is referred to as the Fast
Cell Circulation (FCc).

In case II ($Pr=0.35$) (see figure \ref{fig:Group_picture1} II), the buoyancy current
detaches from the boundary and meanders with a wavelength of about half the cavity
width (figure \ref{fig:Group_picture1}(a) II). (Note that for the case $Pr=0.1$ shown
above an increase in width of the cavity would also result in a detachment of the
buoyancy current). This detachment causes the radiation of internal waves into the
interior (see phase plots in figure \ref{fig:Group_picture1}(c) II) that have a
dominant vertical mode with $\omega / N_c$ close to 1.0, leading to quasi vertical
iso-phase lines. Vorticity perturbations (figure \ref{fig:Group_picture1}(d) II) show a
maximum shear in the detached buoyancy current. The perturbed flow (figure
\ref{fig:Group_picture2}(a,b) II) consists of two large cell structures with a dominant
vertical transport, and large oscillations in the density profile revealing the
presence of internal waves. In contrast to the FCc case above, energy of the detached
buoyancy currents is dispersed into internal wave motions such that the oscillation
frequency is relatively small or 'slow'. This regime is referred to as Slow Circulation
Cells (SCc).

In case III, $Pr=0.53$, (see figures \ref{fig:Group_picture1} III) the flow pattern
changes from a convective flow to a horizontal exchange flow with a shift in direction
at mid-height (figure \ref{fig:Group_picture1}(a) III). The buoyancy currents detach
from the horizontal boundary close to the thermal wall. Oscillations are fast,
$\omega/N_c>3$, and internal waves cannot propagate through the stratified interior,
and the phase plot (figure \ref{fig:Group_picture1}(c) III) shows very different length
scales in the buoyancy currents compared to those in the interior. In the absence of
internal waves and a large scale circulation, the top and bottom motions are decoupled.
Thus, the independent growth of the unstable regions leads to different perturbation
amplitudes implying amplitude asymmetry (see figure \ref{fig:Group_picture1}(b) III).
Because of the fast oscillations and the localisation of the instability in the corner
regions this flow is referred to as the Fast Corner flow (FCo).

In case IV, $Pr=0.71$, (see figures \ref{fig:Group_picture1} IV) we recover the case
studied in detail in former studies \citep[see e.g.][and references
therein]{Paolucci1990, LeQuere1998, Xin2006}, and more recently by \cite{Grayer2020}.
Here $\omega / N_c \approx 0.5$ allowing for internal waves in the interior. These
internal waves and the unstable corner regions are part of the same global instability
mode \citep{Xin2006}, as the phase plot in (figure \ref{fig:Group_picture1}(c) IV)
clearly shows. The amplitude of the internal waves in the interior is relatively weak
compared to the amplitude in the corner regions. The perturbed base flow (figures
\ref{fig:Group_picture2}(a,b) IV) shows next to the oscillating corner regions,
recirculating regions in the interior that are slightly flattened by the internal
buoyancy stratification. The absolute value of the perturbation amplitude of the top
and bottom current are identical or very close, which is referred to as symmetric in
amplitude (not to confuse with centro-symmetry). The vorticity perturbations (figures
\ref{fig:Group_picture1}(d) IV) have opposite signs at the top and bottom corner,
revealing that this mode is anti centro-symmetric (see the definition in \S
\ref{sec:diagnostics}), which is the mode that appears for the lowest Rayleigh number
in agreement with \cite{Burroughs2004, Oteski2015}. The centro-symmetric mode appears
for a slightly higher Rayleigh number.

In case V, $Pr=2.8$, the thermal boundary layer thickness is thin and the perturbation
maxima are limited to very small corner regions (figures \ref{fig:Group_picture1}(a-d)
V). The oscillation frequency is again increased (figure \ref{fig:Group_picture1}(c)
V). Heat is diffused relatively slowly for this $Pr$-number so that the temperature
gradients near the top and bottom boundaries are larger than in the interior. The
scaled oscillation frequency near the boundaries is $\omega/N \approx 0.8$ while in the
interior $\omega/N \approx 1$, implying close to evanescent waves in the interior as
can be inferred from their almost vertical propagation direction (figures
\ref{fig:Group_picture1}(c) V). The coupling between the top and bottom currents is
therefore also weak, and there is amplitude symmetry only for some aspect ratio. In the
interior, there is a smooth exchange flow (figures \ref{fig:Group_picture2}(a,b) V).
This regime is referred to as Fast Corner flow with Evanescent internal waves (FCoE).

In case VI, $Pr=4$, a simulation for a single aspect ratio has been conducted, showing
that the oscillation frequency of the instability (here $\omega/N_c \approx 1.8$)
increases further with $Pr$ (see (see figures \ref{fig:Group_picture1}(a-d) VI).
Internal waves emitted by the buoyancy currents near the top and bottom are evanescent
and cannot propagate further into the interior. The two buoyancy currents have again
different perturbation amplitudes and are therefore asymmetric in amplitude (figure
\ref{fig:Group_picture1}(b) VI). The scales of motion in the boundary layer, top and
bottom buoyancy currents as well as the interior are indeed very different (see figure
\ref{fig:Group_picture1}(c) VI), showing two independent unstable regions. As will be
shown below this is a shear instability located at the wall. This regime is referred to
as the Fast instability at the Wall (FW).

\begin{figure}
\centerline{
\includegraphics[width=0.9\textwidth]{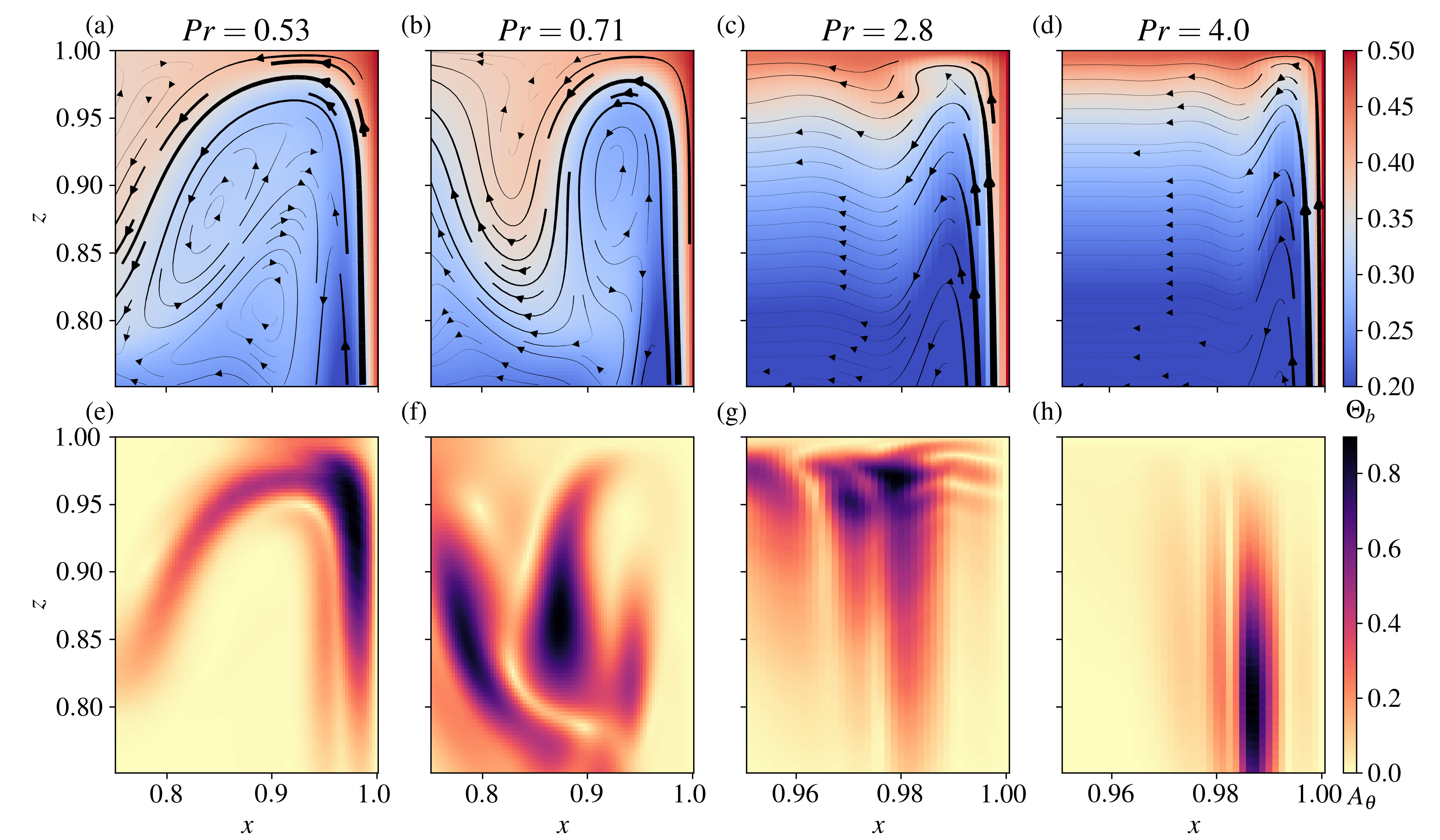}}
\caption{Base flows showing close ups of the streamline -temperature fields of the
corner regions for $A=1$ with maximum amplitude (a-d) for Prandtl numbers from 0.53 to
4.0, and (e-h) the corresponding amplitude fields. Note that zooms and aspect ratios in
(a,b; e,f) and (c,d; g,h) are chosen differently for visualisation purposes.}
\label{fig:zoom}
\end{figure}

\subsubsection{Corner and boundary layer flows}
When zooming in the corner regions (figure \ref{fig:zoom}) substantial flow changes can
be noticed for different $Pr$.  For $Pr=0.53$, the detachment and position of $L_h$
(see figure \ref{fig:sketch}b)) is small. Rolling vortex billows emerging from shear
instability move to the corner (see supplementary movies) leading a maximum of the
instability in the boundary layer near the corner (see figure \ref{fig:zoom}(a,e)). For
$Pr=0.71$, the instability has its maxima in the corner in the detached buoyancy
current, where the main mixing occurs (see figures \ref{fig:zoom}(b,f)). For $Pr=2.8$,
the buoyancy current hardly detaches, and the flow is characterised by the presence of
a downward jet close to the wall (at $0.98$ in figure \ref{fig:zoom}(c,g)). The
instability is located inside the detached buoyancy current and close to the boundary
(figure \ref{fig:zoom}(g)).

For $Pr=4$, the boundary layer detaches right in the corner region where it also
results in a downward return flow (figure \ref{fig:zoom}(d)). The temperature gradient
has its maximum very near to the wall. Outside the boundary layer, the strong shear
between the upward boundary layer motion and the downward return flow with rotary
motions similar as that for Kelvin-Helmholtz type instabilities (see figures
\ref{fig:zoom}(d,h), and supplementary material), suggests a shear instability
\citep[see also][]{Janssen1995, Xu2008}. In the interior, there is a horizontal
exchange flow from right to left above mid depth, and below rom left to right (see
figures \ref{fig:Group_picture1} with $Pr \geq 0.53$). This exchange flow becomes
dominant with increasing $Pr$ number.

The phase plots in figures \ref{fig:Group_picture1}(c)V-VI, ($Pr=2.8$ and $Pr=4.0$)
show a large difference in scale between the boundary layers and the interior. Internal
waves in the interior emerge from the buoyancy currents at the top and bottom
boundaries, and are related to up- and down-ward motions of entrainment and
detrainment, as shown in figures \ref{fig:zoom}(c,d and g,h). Comparing these latter
plots showing the perturbation amplitude (figures \ref{fig:zoom}(g,h)), one notices
that the downward flow for $Pr=4.0$ goes to mid-depth, and for $Pr=2.8$ to only about
half that distance. In contrast ro the case $Pr=2.8$, for $Pr=4$ the instability is
located at the boundary where it has its maximum amplitude.



\subsection{Instabilities and Regime diagram}
\noindent
\begin{figure}
\centerline{\includegraphics[width=0.58\textwidth]{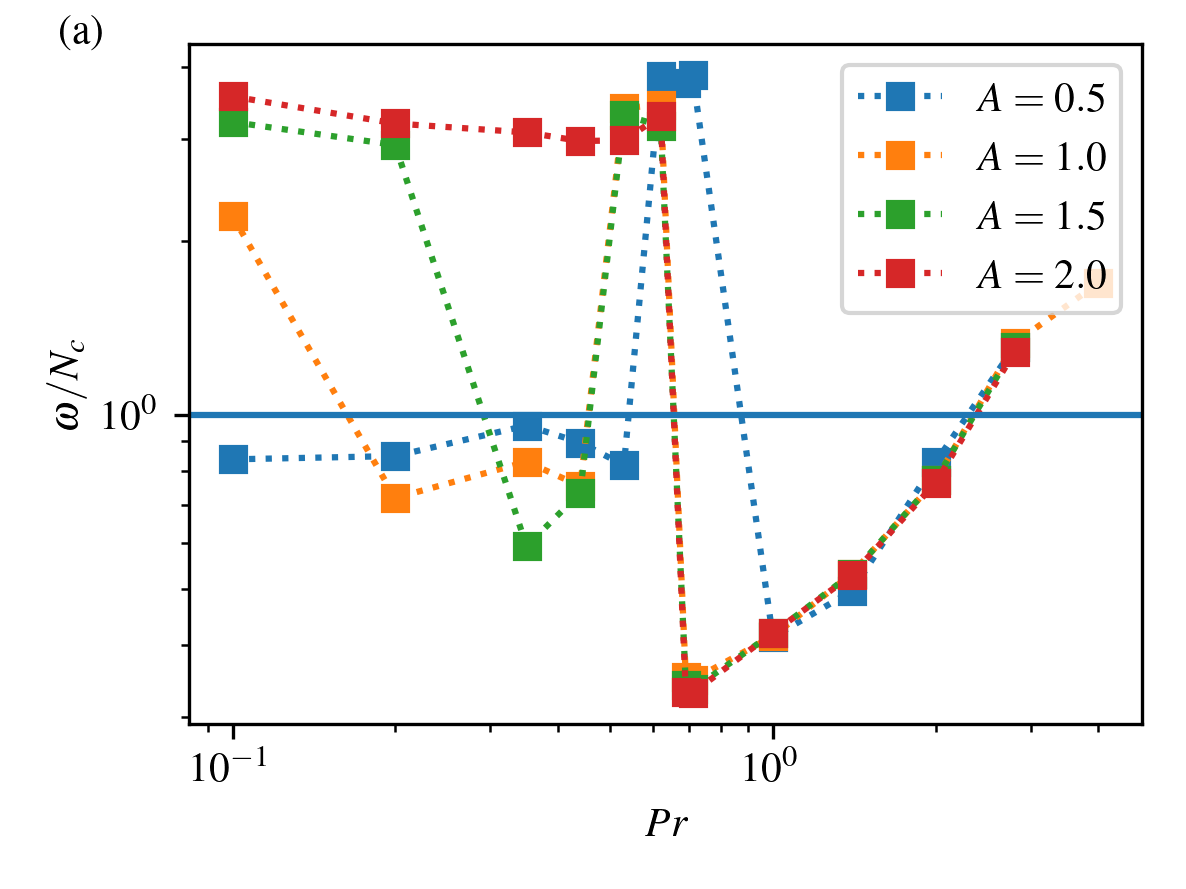}}
\vspace{-3mm}
\centerline{\includegraphics[width=0.58\textwidth]{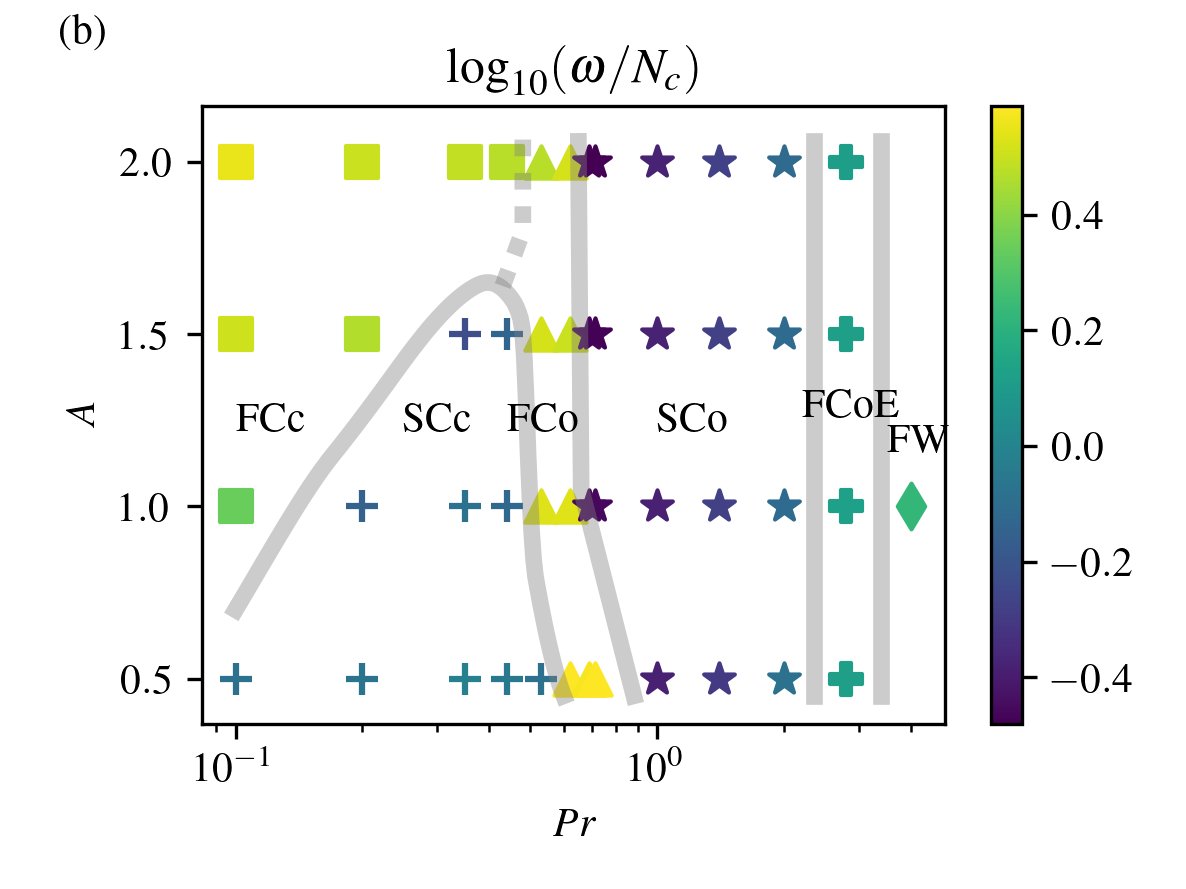}}
\vspace{-3mm}
\centerline{\includegraphics[width=0.58\textwidth]{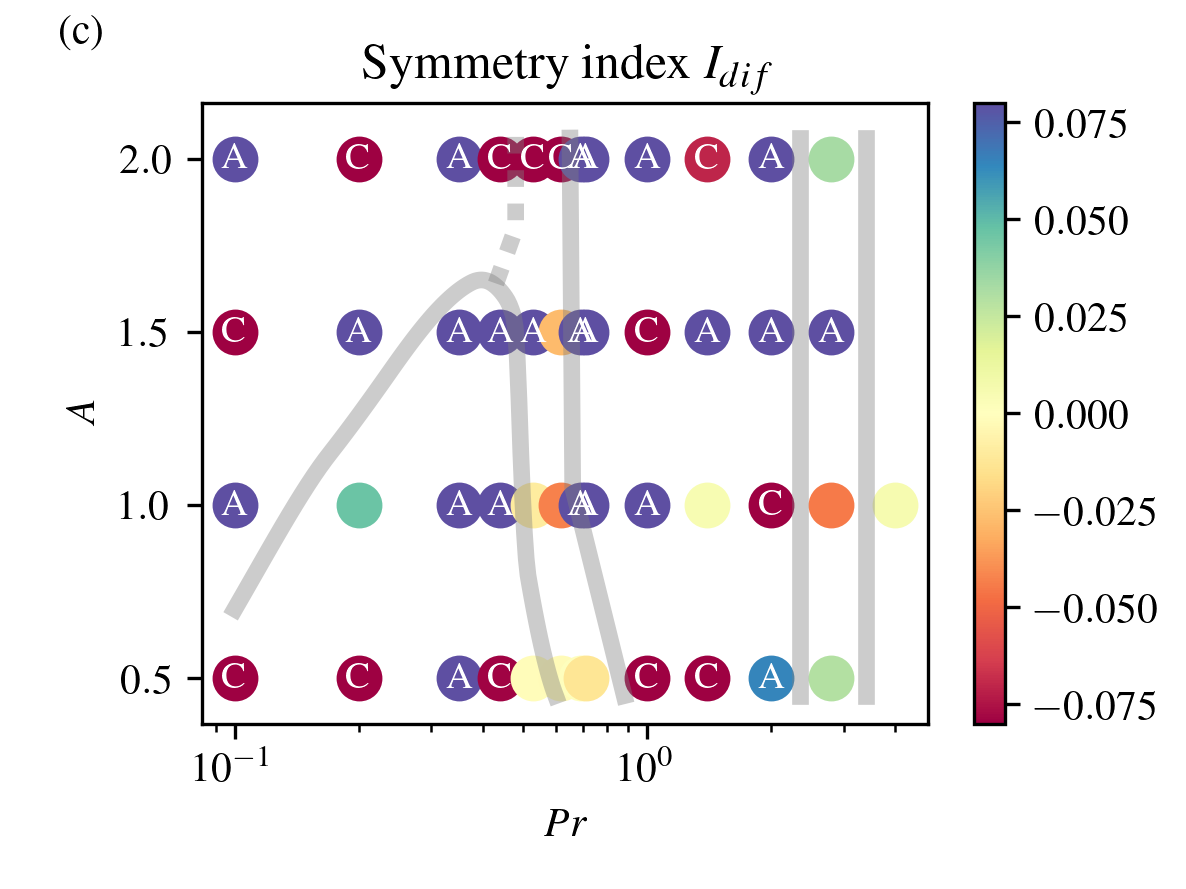}}
\caption{(a) Maximum normalised oscillation frequency ($\omega/N_c$) and (b) regime
diagram in space set by $Pr$ number and aspect ratio $A$, with the color indicating the
logarithmic value of the normalised oscillation frequency $\omega/N_c$. Regime names
are with F for Fast when $\omega/N_c>1$ and S for slow when $\omega/N_c<1$,
$\blacksquare$ Fast Circulation Cells (FCc); {\Large +} Slow Circulation Cells;
$\blacktriangle$ Fast Corner flow (FCo); $\bigstar$ Slow Corner flow (SCo); {\bf{\Large
+}} Fast Corner Evanescent waves (FCoE), and $\blacklozenge$ Fast instability at the
wall (FW). (c) Diagram showing the symmetry index $I_{dif}$ (\ref{centro}), with the
extremes (blue and red) corresponding to symmetric cases (anti centro-symmetric --
letter A -- and centro-symmetric -- letter C --, respectively), and yellow for
amplitude asymmetry and no coupling between top and bottom regions.}
\label{fig:regimes_A_vs_Pr}
\end{figure}
Figure \ref{fig:regimes_A_vs_Pr}(a) presents the normalised oscillation frequency of
the leading linear mode with respect to $Pr$ for the four different values of $A$. Here
$N_c$ varies with $Pr$ (see figure \ref{fig:Lh_vs_Pr}) but  the main variation is in
$\omega$. The sharp jump between at $Pr \approx 0.55 \pm 0.15$ separates two different
regimes, one for small $Pr$ and one for large $Pr$, which can be deduced from scaling
arguments \citep[see][]{Gill1966}. From the stationary heat equation one can deduce
that the velocity along the boundary scales as $\kappa/\ell$ when $\ell$ is the length
scale of the boundary layer. When introducing this scaling into the vorticity equation,
with $\xi=\nabla \times{\bf v_b}$ we obtain for the convective term $ {\bf v_b} \cdot
\nabla \xi \sim \kappa^2 / \ell^3$ and the diffusive term $\nu \nabla^2 \xi \sim \nu
\kappa / \ell^3$. The ratio of the diffusive term over the convective term is equal to
$Pr=\nu / \kappa$.

For small Prandtl number the instability is thus convectively driven with large cell
circulations (see also figure \ref{fig:Group_picture2}). In figure
\ref{fig:regimes_A_vs_Pr}(a) ($Pr \lessapprox 0.5$) the flow is characterised by a cell
circulation that is affected by the cavity aspect-ratio $A$. For small cavity widths
(large $A$), the buoyancy current remains attached to the horizontal boundary, and
reinforces the motion in the thermal boundary layer at the wall,  resulting in a fast
motion with a high frequency of oscillation (see e.g. red squares in figure
\ref{fig:Lh_vs_Pr}(a)). For large cavity widths (small $A$), the buoyancy currents
detach from the horizontal boundaries, and radiate internal waves into the interior.
The cell circulation is therefore weakened, resulting in a relatively slow motion with
a low frequency of oscillation (see e.g. blue squares figure \ref{fig:Lh_vs_Pr}(a)),
and thus a smaller $Pr$-number is needed to increase $\omega/N_c$.

For large Prandtl numbers the diffusion term is larger than the convective term from
which it was concluded that the instability is buoyancy driven
\citep[see][]{McBain2007}. But the detrainment and entrainment motions and the related
return flows that are due to these horizontal boundaries cannot be neglected.
\cite{Janssen1995} (for different $Pr$ numbers, 0.25, 0.71 and 2), and
\cite{Yahata1999} (for $Pr=0.71$) suggested this to be a shear instability with a
change in instability from shear driven for $A >3.65$, to internal wave driven for
$A<3.41$. For the present range in $Pr$ numbers, Figure \ref{fig:regimes_A_vs_Pr}(a)
shows strong variations around $Pr=0.55 \pm 0.15$ before the general trend to thinner
boundaries and larger $\omega/N_c$ values for $Pr \gtrapprox 1$.

The regime diagram in Figure \ref{fig:regimes_A_vs_Pr}(b) shows the oscillation
frequency $\omega/N$ represented by the colour as a function of the $Pr$ number and
aspect ratio $A$, with the symbols indicating the different regimes discussed above.
For $Pr \lessapprox 0.5$ the above reasoning with a dominating cell-driven motion is
refrained in this diagram. But for $Pr > 0.5$, not less than 4 regimes appear due to
the changing influences of internal waves  (as indicated in figure
\ref{fig:regimes_A_vs_Pr}(b) for $Pr>0.5$) and decreasing thickness of the boundary
layer at the wall with $Pr$. In regime FCo there are no waves and no coupling; in SCo
for the majority of cases there are waves and the top and bottom region couple. For
even larger $Pr$ numbers internal waves weaken, and the thermal boundary layers are
thinner (FCoE). For $Pr = 4.0$ (regime FW) there are no waves and no coupling. In
contrast to the regime FCo, the instability occurs at the boundary. Since for larger
$Pr$, the lateral boundary layers will be even closer to the walls, one may speculate
that this regime will not further change. Simulations were tested for aspect ratio,
$A=6$ and $Pr=0.7$ as in \cite{Xin2006} showing again the FCc-regime. In view of the
mentioned increase of the boundary layer thickness with height, and the increasing
inertia of the buoyancy current for taller cavities and thus larger $L_h$, the regime
FCo most likely disappears for larger aspect ratio $A$.

Figure \ref{fig:regimes_A_vs_Pr}(c) shows the amplitude asymmetry according to the
definition of equation (\ref{centro}) as a function of $Pr$ and $A$. In the FCo-regime
due to the absence of internal waves and cell circulation the flow is asymmetric in
amplitude. For larger $Pr$ (regime SCo) waves are present, and generally (except for
$A=1$, $Pr=1.4$) couple top and bottom regions. Not all flows with internal waves are
found to be symmetric in amplitude. Wave patterns in the interior vary considerably
depending on aspect ratio of the cavity and the location of the detachment of the
buoyancy currents, and some do not allow for coupling. This may explain the isolated
points of asymmetry in amplitude (orange, green and yellow) for regimes with internal
waves (SCc, SCo in figure \ref{fig:regimes_A_vs_Pr}(b)). In the regime FCoE ($Pr =
2.8$) internal waves are close to evanescent, so that the coupling is generally weak in
this regime, causing asymmetry in the amplitude for most cases, and complete asymmetry
in the regime FW ($Pr=4$). Regime FCc is symmetric in amplitude due to the coupling by
the cell circulation, and SCc due to internal waves except for one case ($A=1$ and
$Pr=0.2$). For $Pr \gtrapprox 0.5$ the cell circulation was found to increase in
strength for larger aspect ratio $A>1$, and appeared to cause amplitude symmetry also
in the regime FCo, thus showing some overlap between the regimes (dashed line in
figures \ref{fig:regimes_A_vs_Pr}(b,c)). This suggest that for higher aspect ratio $A
\geq 2$ and $Pr < 0.6$ the flow is also dominated by convectively driven cells.

Figure \ref{fig:regimes_A_vs_Pr}(c) also shows anti centro- and centro-symmetry
(letters A and C, respectively). These results are coherent with former results (see
e.g. \cite{Burroughs2004, Oteski2015, Xin2006}). However, no systematic variation with
the regime diagram and amplitude asymmetry was found in this context. It is expected
that more pertinent results could be found with a continuous parameter variation method
\citep[see e.g.][]{Lyubimova.2009}. This is left open for further research.

Figures \ref{fig:regimes_A_vs_Pr}(a,b) suggest that there is a transition in
instability in the regime $Pr =0.55 \pm 0.15$. The increase of both, the shear and the
density gradient between the detached buoyancy currents and the motion in the boundary
layer with $Pr$, make it hard, however, to determine when exactly the instability is
buoyancy or shear driven. A quantitative comparison is needed to determine the precise
nature of the instability for each $Pr$ number. This is undertaken in a separate study.

Outside the space of this regime diagram, for aspect ratio greater than 3 or 4, the
boundary layer will become unstable with, for large $Ra$, the detachment of vortices
disturbing the internal stratification and generating internal waves.
\citep[][]{Xin1995}. In the limit of very large $Pr$ and large $A$, the onset of local
cells was found in the core, the number of the cells being set by the scale of the
instability in the boundary layer \cite[see][and references
therein]{Daniels1985,Daniels1987}

\begin{figure}
\centerline{
\includegraphics[width=0.49\textwidth]{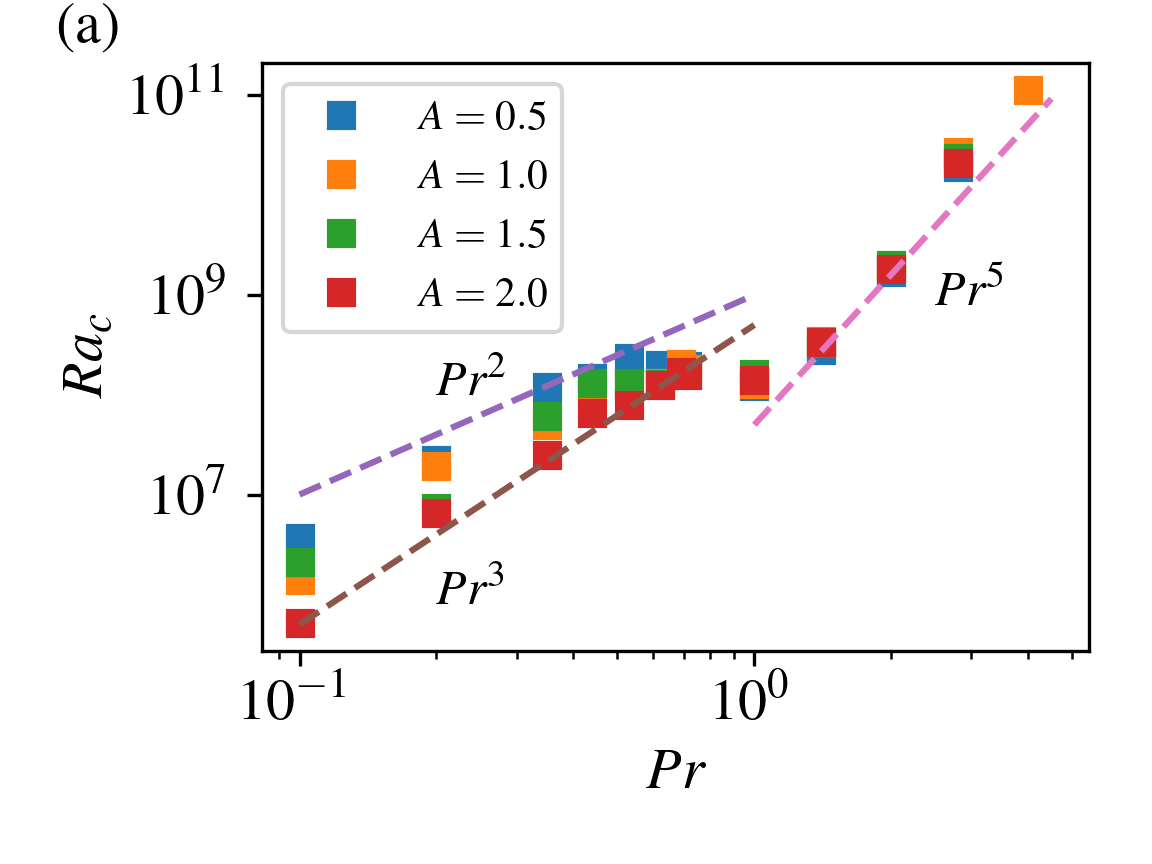}
\includegraphics[width=0.49\textwidth]{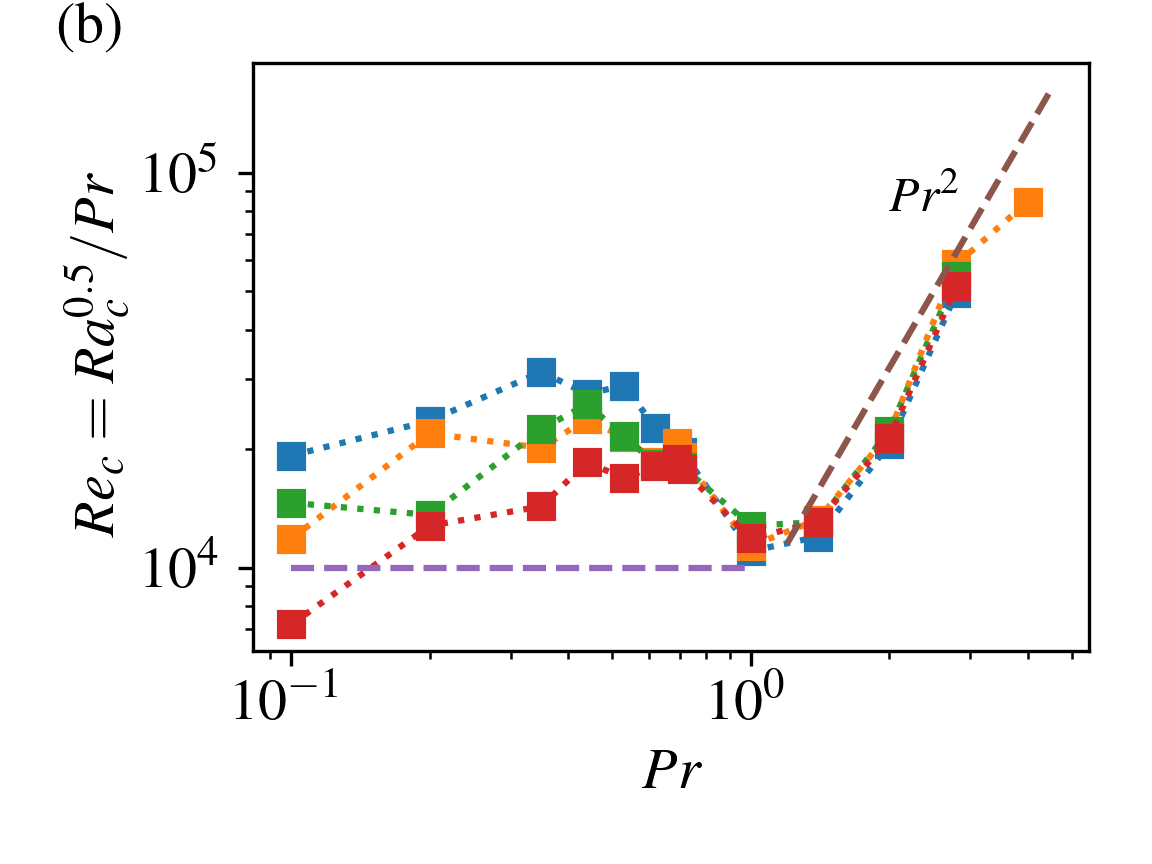}
}
\caption{Critical Rayleigh number as a function of Prandtl number and (b) the Reynolds
number derived from the critical Rayleigh number and Prandtl numbet agains Prandtl
number for different aspect ratio $A$ (see legend). The dashed lines represent the
power laws discussed in the text.} \label{fig:RacGrcRec_vs_Pr}
\end{figure}

Figures \ref{fig:RacGrcRec_vs_Pr}(a) and (b) show the variation of the critical
Rayleigh number and Reynolds number as a function of $Pr$, respectively. There is a
clear transition between the two regimes, with $Ra_c \sim Pr^2$ for $Pr<1$ and
$Ra_c\sim Pr^5$ for $Pr>1$. In between, (i.e. for $0.5<Pr \lessapprox 2$), there is an
intermediate region where internal waves are present and play a role for the dynamics.
In both figures (\ref{fig:RacGrcRec_vs_Pr} a and b), for $Pr>1$ the aspect ratio does
not affect the results. Finally, it is noteworthy that the Reynolds number defined as
$Re_c=Ra_c^{0.5}/Pr$ could serve as an appropriate critical value for the onset of
instability since it varies just between $(2\pm1) 10^4$ for $Pr<1$, and for $Pr>1$,
$Re_c=10^4 Pr^2$, whereas the Rayleigh number varies over 6 decades (see figure
\ref{fig:RacGrcRec_vs_Pr}(b)).  This Reynolds number compares well with a Reynolds
number based on the maximum velocity in the cavity and the appropriate typical length
scale.


\section{Conclusions and discussion}
\label{sec:conclusions}
The present investigation shows that there is a large variation in flow regimes
depending on the Prandtl number and aspect ratio as is represented in the regime
diagram of figure \ref{fig:regimes_A_vs_Pr}(b). The role of the detachment in the
corner regions and the internal waves on the dynamics is investigated.

The regimes depend on the variation in the detachment of the buoyancy current with
Prandtl number and its effect on the circulation inside the cavity, and on the other
hand, the presence of internal waves.  When there is no detachment (for $Pr<0.5$) and
the buoyancy current is limited by the horizontal extend of the cavity, a cell
circulation develops. The instability mode is global, and top and bottom motions have
amplitude symmetry, i.e. they have same (absolute) perturbation amplitude.

When there is detachment of the buoyancy current, there is a local circulation that is
determined by the dynamics of the buoyancy current. The coupling between top and bottom
motions, and therewith their symmetry in amplitude then depends on the presence of
internal waves. Generally, in the absence of internal waves and large-scale
circulation, the perturbation amplitudes in the corner region differ and the flow is
asymmetric. However, some exceptions occur, and not all internal wave patterns allow
for this coupling.

The critical Rayleigh number depends on the Prandtl number, with two regimes, one for
$Pr<1$ with some variation due to aspect ratio and $Ra_c \sim Pr^2$ , and one for
$1<Pr<4$ where the aspect ratio does not affect the results and $Ra_c \sim Pr^5$. As
mentioned above, this can equally be expressed in terms of the Reynolds number. For our
extreme value of $Pr=4$, \cite{Janssen1995} find a critical Rayleigh number $Ra_c=2.5
\cdot 10^{10}$, while for $Pr=10$ \cite{Wang2021} find a critical Rayleigh number of
$Ra_c = 6 \cdot 10^{10}$, showing a downward tendency. This subject is open to further
research. On the other hand, for very small Prandtl numbers, \cite{Gelfgat1999} found
critical $Ra$ numbers of order O($10^{-4}$) so that on this side, a further decrease in
critical Rayleigh number can be expected with higher critical Rayleigh numbers for the
smaller aspect ratio $A$.

Some of the regimes represented in figure \ref{fig:regimes_A_vs_Pr}(b) have been
observed in former investigations. For air-filled cavities of aspect ratio of 6 and 7
e.g. the FCc regime was also found in \cite{Xin2006}, whereas the SCo regime is well
studied in e.g. \cite{LeQuere1998} and \cite{Oteski2015}, but no detailed information
has been found about the instabilities of the FCoE- and FW-regimes. Most remarkable is
the FCo-regime with its drastic increase in oscillation frequency of a factor of 10 for
Prandtl numbers in the range of $Pr=0.55 \pm 0.15$. This regime has not been shown
before. It most likely disappears for larger aspect ratio, with for $A \leq 2$ and
approximately $Pr < 0.55$ the flow being dominated by convectively driven cells.

For $0.4 \leq Pr \leq 0.7$, the shear and the temperature gradient between the corner
region and upward motion near the wall increase, and the corner region oscillates with
internal waves in the interior. There is uncertainty about the origin of the
instability mechanism, and the roles of shear and internal waves. For $Pr>2.8$ the
region of instability changes again and occurs between the downward motion in the
detached corner flow and the upward motion at the wall at some distance from the
boundary. Also here the temperature gradient is large. Therefore, in order to determine
the origin of the instability, a comparison of the individual terms in the momentum
equations, as well as plots for Rayleigh criterion, centrifugal instability, and
Richardson number is needed. Since this is a rather elaborate effort, it will be
presented elsewhere.

This investigation is limited to a single mode, obtained for the lowest critical
$Ra$-number. When increasing $Ra$, bifurcations with other modes may appear, as shown
by \cite{Oteski2015} for $Pr=0.7$ for air, and in a three dimensional box by
\cite[e.g.][]{Gelfgat2017}. In view of former results obtained with DNS
\cite[see][]{Trias2007}, one may nevertheless expect that the present results will
provide also a good guide line for the three dimensional case, as long as the Rayleigh
number is small ($Ra< 10^{10}$). Preliminary tests in the present research have shown
that three-dimensional instabilities are absent as long as the cavity depth is about
10\% of the cavity-height (i.e. $0.1 \times H$). Therefore we expect that the present
regimes have their footprint also in a quasi three-dimensional environment.

\backsection[Supplementary data]{\label{Movie1}Supplementary material and movies are
available at \\https://doi.org/10.1017/jfm.2019...}

\backsection[Acknowledgments]{
The authors thank the anonymous referees for their helpful comments. Further, they
acknowledge Olivier De-Marchi, Gabriel Moreau and Cyrille Bonamy of the LEGI's
informatics team for their support. A CC-BY public copyright license has been applied by the authors to the
present document and will be applied to all subsequent versions up to the Author
Accepted Manuscript arising from this submission, in accordance with the grant's open
access conditions.}

\backsection[Funding] {This project was funded by the project
LEFE/IMAGO-2019 contract COSTRIO. AK acknowledges the finance of his PhD thesis from
the school STEP of the University Grenoble Alpes. Part of this work was performed using
resources provided by \href{https://www.cines.fr/}{CINES} under GENCI allocation number
A0120107567.}

\backsection[Declaration of Interests]{The authors report no conflict of interest.}

\backsection[Data availability statement.]{ The data that support the findings of this study are openly available in Zenodo at https://zenodo.org/records/7827872, reference number 7827872.}

\backsection[Author ORCID's]{\\A. Khoubani https://orcid.org/0000-0002-0295-5308; \\ A. V. Mohanan https://orcid.org/0000-0002-2979-6327; \\P. Augier https://orcid.org/0000-0001-9481-4459; \\J.-B. Fl\'or https://orcid.org/0000-0002-7114-2263.}

\begin{table}
\appendix\section{}
\begin{center}
\begin{tabular}{rrrrrr l rrrrrr}
\hline
  $A$ &  $Pr$ &    $Ra (10^7)$ &  $nx$ &  $ny$ &  $\;\;$ &  $ A $ &  $Pr$ &    $Ra(10^7)$ & $nx$ &  $ny$\\
 \hline
 0.5 & 0.1 & 0.369&  80 &  40 &    & 1  & 0.1 & 0.1406 &  40 &  40 \\
 0.5 & 0.2 & 2.2 &  80 &  40 &     & 1  & 0.2 & 2.605&  40 &  40 \\
 0.5 & 0.35 & 12.8 &  80 &  40 &     & 1  & 0.35 & 4.939 &  40 &  40 \\
 0.5 & 0.44 & 16.5 &  88 &  44 &     & 1  & 0.44 & 10.962 &  44 &  44 \\
 0.5 & 0.53 & 24 &  88 &  44 &     & 1  & 0.53 & 14.82 &  44 &  44 \\
 0.5 & 0.62 & 24.1 &  88 &  44 &     & 1  & 0.62 & 13.26 &  44 &  44 \\
 0.5 & 0.69 & 19.85 &  88 &  44 &     & 1  & 0.69 & 20.40 &  44 &  44 \\
 0.5 & 0.71 & 23.75 &  88 &  44 &     & 1  & 0.71 & 18.35 &  44 &  44 \\
 0.5 & 1  & 12.13 &  96 &  48 &     & 1  & 1  & 12.88&  48 &  48 \\
 0.5 & 1.4 & 28.15 & 104 &  52 &     & 1  & 1.4 & 35.15&  52 &  52 \\
 0.5 & 2  & 310 & 160 &  80 &     & 1  & 2  & 2000  &  80 &  80 \\
 0.5 & 2.8 & 2880 & 200 & 100 &    & 1  & 2.8 & 2800& 100 & 100 \\
    &    &         &     &    &     & 1  & 4   & 11950  & 100 & 100 \\
\hline
\end{tabular}
\end{center}
\vspace{1.5cm}
\begin{center}
\begin{tabular}{rrrrrr l rrrrrr}
\hline
  $A$ &  $Pr$ &    $Ra (10^7)$ &  $nx$ &  $ny$ &  $\;\;$ &  $ A $ &  $Pr$ &    $Ra(10^7)$ & $nx$ &  $ny$\\
\hline
 1.5 & 0.1 &   0.213 &  27 &  40 &     & 2  & 0.1 &   0.594200    &  20 &  40    \\
 1.5 & 0.2 &   0.739 &  27 &  40 &     & 2  & 0.2 &   0.6408 &  20 &  40 \\
 1.5 & 0.35 &   6.19 &  27 &  40 &     & 2  & 0.35 &   2.6788 &  20 &  40 \\
 1.5 & 0.44 &   13.2  &  33 &  50 &     & 2  & 0.44 &   6.7 &  28 &  56 \\
 1.5 & 0.53 &   13.18  &  33 &  50 &     & 2  & 0.53 &   7.95 &  25 &  50 \\
 1.5 & 0.62 &   12.135  &  33 &  50 &     & 2  & 0.62 &   12.63 &  25 &  50 \\
 1.5 & 0.69 &   18.26  &  33 &  50 &     & 2  & 0.69 &   17 &  25 &  50 \\
 1.5 & 0.71 &   16.325 &  33 &  50 &     & 2  & 0.71 &   15.87 &  25 &  50 \\
 1.5 & 1  &   158 &  37 &  56 &     & 2  & 1  &   14.125 &  28 &  56 \\
 1.5 & 1.4 &   34.1 &  40 &  60 &     & 2  & 1.4 &   34.15 &  30 &  60 \\
 1.5 & 2  &    197.2 &  53 &  80 &     & 2  & 2  &   181.25  &  40 &  80 \\
 1.5 & 2.8 &   2600 &  80 & 120 &     & 2  & 2.8 &   2500 &  60 & 120 \\
\hline
\end{tabular}
\caption{Input parameters for the linear simulations. $nx$ and $ny$ are number of
elements in x and y directions, respectively, and the polynomial order $O$ is for all
cases 7 so that the number of grid points is $(O+1)^2 \times nx \times ny$.}
\end{center}
\label{appendix}
\end{table}
\bibliographystyle{jfm}
\bibliography{biblio}
\end{document}